\documentclass{article}

\usepackage{arxiv}
\IfFileExists{NJDnatbib.sty}{\usepackage[numbers,super]{NJDnatbib}}{\usepackage[numbers,super]{natbib}}
\usepackage{booktabs}
\usepackage{enumitem}
\usepackage{nicefrac}
\usepackage[utf8]{inputenc}
\usepackage[T1]{fontenc}
\usepackage{tikz}
\usepackage{amsmath}
\usepackage{amsthm}
\usepackage{amssymb}
\usepackage{graphicx}
\usepackage{multirow}
\usepackage{xcolor}
\usepackage{comment}
\usepackage{float}
\usepackage{hyperref}       
\usepackage{url}            
\usepackage{booktabs}       
\usepackage{amsfonts}       
\usepackage{nicefrac}       
\usepackage{microtype}      
\usepackage{lipsum}		    

\usetikzlibrary{shapes.geometric,arrows,fit,matrix,positioning}
\tikzstyle{stepx} = [rectangle, rounded corners, minimum width=3cm, minimum height=1cm,text centered, draw=black, fill=red!30]
\tikzstyle{arrow} = [thick,->,>=stealth,text width=8.5cm]

\usepackage{todonotes}
\usepackage{rotating}

\newcommand{\SR}[1]{}
\newcommand{\grey}[1]{\textcolor{gray}{#1}}

\usepackage{todonotes}
\usepackage{rotating}

\title{Empirical Prior Distributions for Bayesian Meta-Analyses of Binary and Time to Event Outcomes}

\author{
         František Bartoš               \\
	Department of Psychology       \\
	University of Amsterdam        \\
	Noord-Holland, The Netherlands \\
	\And
 	Willem M. Otte \\
	Department of Pediatric Neurology,\\ UMC Utrecht Brain Center\\
	University Medical Center Utrecht\\
	Utrecht, The Netherlands \\
	\And
	Quentin F. Gronau              \\
	Department of Psychology       \\
	University of Amsterdam        \\
	Noord-Holland, The Netherlands \\
	\And
	Bram Timmers                   \\
	Department of Psychology       \\
	University of Amsterdam        \\
	Noord-Holland, The Netherlands \\
	\And
	Alexander Ly                   \\
	Department of Psychology       \\
	University of Amsterdam        \\
	Noord-Holland, The Netherlands \\
	\& \\
	Centrum Wiskunde \& Informatica, Amsterdam \\
	Noord-Holland, The Netherlands \\
	\And
	Eric-Jan Wagenmakers           \\
	Department of Psychology       \\
	University of Amsterdam        \\
	Noord-Holland, The Netherlands \\
	}

\renewcommand{\shorttitle}{Empirical Prior Distributions for Bayesian Meta-Analyses}

\begin{document}
\maketitle

\begin{abstract}
Bayesian model-averaged meta-analysis allows quantification of evidence for both treatment effectiveness $\mu$ and across-study heterogeneity $\tau$. We use the Cochrane Database of Systematic Reviews to develop discipline-wide empirical prior distributions for $\mu$ and $\tau$ for meta-analyses of binary and time-to-event clinical trial outcomes. First, we use 50\% of the database to estimate parameters of different required parametric families. Second, we use the remaining 50\% of the database to select the best-performing parametric families and explore essential assumptions about the presence or absence of the treatment effectiveness and across-study heterogeneity in real data. We find that most meta-analyses of binary outcomes are more consistent with the absence of the meta-analytic effect or heterogeneity while meta-analyses of time-to-event outcomes are more consistent with the presence of the meta-analytic effect or heterogeneity. Finally, we use the complete database - with close to half a million trial outcomes - to propose specific empirical prior distributions, both for the field in general and for specific medical subdisciplines. An example from acute respiratory infections demonstrates how the proposed prior distributions can be used to conduct a Bayesian model-averaged meta-analysis in the open-source software \texttt{R} and JASP.
\end{abstract}


\newpage
\section{Introduction}





Clinical trials are essential for testing whether novel therapeutic treatments and inverventions are indeed benificial and not harmful.\cite{friedman2015fundamentals} Before novel therapeutic interventions can be implemented, multiple related clinical trials are required to accumulate evidence for the presence of a beneficial treatment effect. This process is usually achieved by meta-analytical techniques that allow researchers to pool estimates from individual clinical trials and obtain an aggregated estimate of the treatment effectiveness and its uncertainty.\cite{borenstein2009introduction} If the overall uncertainty is too large, additional data collection might be required to obtain a more precise estimate.

Aggregating data from multiple clinical trials poses a non-trivial analysis problem as the selection of a meta-analysis model affects the treatment effect point estimate and uncertainty jointly.\cite{brockwell2001comparison,berkey1995random} Uncertainty may originate from heterogeneity within and between trial data. For example, a fixed-effects meta-analysis model for the joint analysis of multiple clinical trials assumes no between-trial uncertainty, whereas a random-effects model takes both within and between-trial heterogeneity into account.\cite{borenstein2009introduction} In the case with few clinical trials, it is difficult to establish whether a fixed-effects or random-effects model is most appropriate for the data at hand. Furthermore, the traditional approaches struggle with obtaining a reliable between-trial heterogeneity estimate.\cite{brockwell2001comparison, inthout2014hartung, gonnermann2015no} The required choice between fixed or random meta-analysis model is of limited interest to the researcher pooling multiple clinical trials, as the concern is with obtaining the pooled effect estimate rather than the most appropriate model.

In a previous article,\cite{bartos2021bayesian} we introduced Bayesian model-averaged (BMA) meta-analysis\cite{gronau2021primer, gronau2017powerpose} as a coherent way of combining meta-analytic inference across fixed- and random-effect models, and developed empirical prior distributions for the treatment effectiveness $\delta$ and across-study heterogeneity $\tau$ parameters for Cohen's $d$ measured continuous outcomes. The BMA meta-analysis allows researchers to combine inferences based on a series of competing models by taking each of them into account according to their prior predictive performance and draw inferences that incorporate uncertainty about the data-generating process. Furthermore, the empirical prior distributions improve parameter estimates under small sample sizes (also see \cite{williams2018bayesian, higgins2009re, chung2013avoiding,rhodes2016implementing,higgins1996borrowing}) and specify Bayes factors tests for the presence vs. absence of the treatment effectiveness and between-study heterogeneity.\cite{gronau2021primer, kass1995bayes, jeffreys1961theory, hinne2019conceptual}

In this article, we extend the previous continuous-outcome-only work by proposing empirical prior distributions for both the treatment effectiveness and across-study heterogeneity for binary and time-to-event outcomes based on medical data obtained from the Cochrane Database of Systematic Reviews (CDSR). We use the binomial-normal model for log odds ratios and a normal-normal model for log risk ratios, risk differences, and log hazard ratios\footnote{This is also an extension to Pullenayegum\cite{pullenayegum2011informed} and Turner and colleagues\cite{turner2015predictive} who developed prior distributions for the across-study heterogeneity $\tau$ based on an earlier version of the database.} Furthermore, we summarize information about the evidence in favor of the effect and across-study heterogeneity in the CDSR, implement the empirical prior distributions in the open-source software \texttt{R}\cite{R} and JASP,\cite{JASP17, ly2021bayesian} and illustrate the methodology on an acute respiratory infections example.

\section{Meta-Analysis of Binary and Time to Event Outcomes}

\subsection{Effect size measures}
We examine meta-analytic models of binary and time-to-event outcomes (see Bartoš and colleagues\cite{bartos2021bayesian} for treatment of continuous outcomes). While both outcome types can be, in essence, addressed in the same way as continuous data, there are additional specificities when dealing with binary outcomes.

\begin{table}[h]
    \centering
    \begin{tabular}{l|cc|c}
            &  Events & Non-Events & N     \\
    \hline
    Group 1 &  a      & b          & $\text{n}_1$ \\
    Group 2 &  c      & d          & $\text{n}_2$ \\
    \hline
    \end{tabular}
    \caption{Characteristics of outcomes of a binary study.}
    \label{tab:binary_outcomes}
\end{table}

By binary outcomes, we refer to studies whose endpoint of interest is either a presence or absence of an event (e.g., death or recurrence). Consequently, results of individual trials can be summarized by a $2 \times 2$ table such as Table~\ref{tab:binary_outcomes}. The rows in Table~\ref{tab:binary_outcomes} correspond to the study groups (e.g., treatment vs control arm), the first column denotes the number of observed events, the second row denotes the number of event-free observations, and the third column denotes the number of observation in each group. 

There are multiple ways to quantify the differences between the two groups (i.e., effect sizes), the most common being odds ratios (OR), risk ratios (RR), and risk differences (RD). Each of the measures has its merits and the selection of the most appropriate measure needs to be based on clinical considerations. The most popular choices are the relative effect size measures (i.e., OR and RR) which are less sensitive to the baseline event rate; however, it has been argued that risk differences are better suited for convening the clinical impact.\cite{holmberg2020estimating, schechtman2002odds, doi2020controversy}

\begin{table}[h]
    \centering
    \begin{tabular}{l|rr}
            &  Effect Size & Standard Error    \\
    \hline
    log OR &  $\text{log}\left(\frac{\nicefrac{a}{b}}{\nicefrac{c}{d}}\right)$      & $\sqrt{\frac{1}{a} + \frac{1}{b} + \frac{1}{c} + \frac{1}{d}}$     \\
    log RR &  $\text{log}\left(\frac{\nicefrac{a}{n_1}}{\nicefrac{c}{n_2}}\right)$ & $\sqrt{\frac{1}{a} - \frac{1}{n_1} + \frac{1}{b} - \frac{1}{n_2}}$ \\
    RD     &  $\frac{a}{n_1} - \frac{c}{n_2}$                           & $\sqrt{\frac{ab}{n_1^3} + \frac{cd}{n_2^3}}$                       \\
    \hline
    \end{tabular}
    \caption{Effect sizes and approximate standard errors of effect size measures of binary outcomes based on formulas from Borenstein and colleagues 2009.\cite{borenstein2009introduction}}
    \label{tab:binary_measures}
\end{table}

OR and RR are ratios (as suggested by the name) -- they are asymmetric and can attain positive values only. Therefore, log OR and log RR are usually used which leads to symmetric and unbounded response variables that can be modeled via a normal likelihood. Table~\ref{tab:binary_measures} summarizes log OR, log RR, and RD and their standard errors. 

Table~\ref{tab:binary_measures} also highlights an additional issue with the usage of log OR and log RR -- both are undefined when the number of events in either of the groups is zero. This happens suprisingly often; $22.7$\% of binary outcomes meta-analyses included in CDSR contain at least one cell with zero events. One way of dealing with the zero-cell issue is continuity corrections\cite{plackett1964continuity} which add a small positive number to cells in Table~\ref{tab:binary_outcomes} or the Mantel-Haenszel method\cite{mantel1959statistical} or binomial-normal models for OR.

For time-to-event outcomes, it is natural to use the log hazard ratios (log HR), which take into account the number of events, the timing of events, and the time until the last follow-up for each trial participant without an event (i.e., right censoring).\cite{tierney2007practical, simmonds2011meta}

\subsection{Normal-normal model}
Subsequently, the log OR, log RR, RD, or log HR, $\text{y}_i$, and their standard errors, $\text{se}_i$, can be modeled using a normal-normal meta-analytic model,
\begin{align}
    \label{mod:normal-normal}
    \text{y}_i &\sim \text{Normal}(\gamma_i, \text{se}_i),\\
    \nonumber
    \gamma_i &\sim \text{Normal}(\mu, \tau),
\end{align}
where Normal($x$, $y$)'' denotes a normal distribution with mean $x$ and standard deviation $y$. The model parameters, $\mu$ and $\tau$ correspond to the mean effect size and between study standard deviation (heterogeneity) and $\gamma_i$ correspond to normally distributed true study effects. If we assume the absence of heterogeneity, i.e., a fixed effect model, $\tau$ is set to zero, and all $\gamma_i$ are equal to $\mu$.

\subsection{Binomial-normal model}
The continuity corrections with a normal-normal meta-analytic model can lead to bias, especially in unbalanced designs.\cite{sweeting2004add} Another, more elegant, solution in the case of log OR is a binomial-normal logistic model that naturally deals with zero events,\cite{gunhan2020random, smith1995bayesian}
\begin{align}
    \label{mod:binomial-normal}
    \text{a}_i &\sim \text{Binomial}(\pi_{1,i}, \text{n}_{1,i}),\\
    \nonumber
    \text{c}_i &\sim \text{Binomial}(\pi_{2,i}, \text{n}_{2,i}),\\
    \nonumber
    \text{logit}(\pi_{1,i}) &= \beta_i + \gamma_i/2,\\
    \nonumber
    \text{logit}(\pi_{2,i}) &= \beta_i - \gamma_i/2,\\
    \nonumber
    \gamma_i &\sim \text{Normal}(\mu, \tau),
\end{align}
where Binomial($\pi$, $n$) denotes a binomial distribution with the probability of an event $\pi$ and the number of observations $n$. The parameter $\beta_i$ correspond to study specific base-rate probabilities (on logistic scale), and parameters $\mu$, $\tau$, and $\gamma_i$ have the same interpretation as in Equation~\ref{mod:normal-normal}.

\section{Bayesian Model-Averaged Meta-Analysis}

When considering the normal-normal or binomial-normal meta-analytic models, we need to specify prior distributions for the $\mu$ and $\tau$ parameters defining the competing hypotheses $\mathcal{H}_\cdot$  (and $\beta_i$ in the case of the binomial-normal). Different prior distribution specifications/assumptions about either the presence or absence of the $\mu$ and $\tau$ parameters lead to four qualitatively different models;
\begin{enumerate}[itemsep=0mm]
    \item the fixed-effect null hypothesis $\mathcal{H}_0^f$  : $ \mu = 0$ , $\tau = 0$,
    \item the fixed-effect alternative hypothesis $\mathcal{H}_{1}^f$  : $ \mu \sim g(\cdot)$ , $\tau = 0$,
    \item the random-effects null hypothesis $\mathcal{H}_0^r$ : $\mu = 0$, $\tau \sim h(\cdot)$,
    \item the random-effects alternative hypothesis $\mathcal{H}_{1}^r$  : $\mu \sim g(\cdot)$ , $\tau \sim h(\cdot)$,
\end{enumerate}
where $g(\cdot)$ denotes a prior distribution on the mean effect size parameter assuming the presence of an effect, and $h(\cdot)$ denotes a prior distribution on the heterogeneity parameter assuming the presence of heterogeneity.

The goals of Bayesian model-averaged meta-analysis (BMA) are twofold: 1) to evaluate the evidence in favor/against the specified models and hypotheses; and 2) to combine parameter estimates from the specified models. BMA provides a coherent way of combining results from multiple competing models (see Gronau and colleagues\cite{gronau2021primer} and Bartoš and colleagues\cite{bartos2021bayesian} for detailed treatment). This is in a stark difference to the classical inference that usually proceeds by selecting one of the models (2 or 4) on either theoretical or quantitative grounds (by the very so often under-powered test for residual heterogeneity) and completely commits to the single selected model.

In short, in BMA, evidence in favor of the effect is quantified via an inclusion Bayes factor which measures the change from prior to posterior inclusion odds for the effect,
\begin{equation}
\label{eq:BF_effect}
\underbrace{\text{BF}_{10}}_{\substack{\text{Inclusion Bayes factor}\\{\text{for effect}}}} 
= \;\;
\underbrace{ \frac{p(\mathcal{H}_{1}^f \mid \text{data}) + p(\mathcal{H}_{1}^r \mid \text{data})}{p(\mathcal{H}_{0}^f \mid \text{data}) + p(\mathcal{H}_{0}^r \mid \text{data})}}_{\substack{\text{Posterior inclusion odds}\\{\text{for effect}}}}
\;\; \bigg/ \;\; 
\underbrace{ \frac{p(\mathcal{H}_{1}^f) + p(\mathcal{H}_{1}^r)}{p(\mathcal{H}_{0}^f) + p(\mathcal{H}_{0}^r)}}_{\substack{\text{Prior inclusion odds}\\{\text{for effect}}}},
\end{equation}
where $p(\mathcal{H}_\cdot)$ denotes prior model probabilities and $p(\mathcal{H}_\cdot \mid \text{data})$ denotes posterior model probabilities. Similarly, the inclusion Bayes factor for heterogeneity is given by the change from prior to posterior inclusion odds for the heterogeneity,
\begin{equation}
\label{eq:BF_heterogeneity}
\underbrace{\text{BF}_{rf}}_{\substack{\text{Inclusion Bayes factor}\\{\text{for heterogeneity}}}} 
= \;\;
\underbrace{ \frac{p(\mathcal{H}_{0}^r \mid \text{data}) + p(\mathcal{H}_{1}^r \mid \text{data})}{p(\mathcal{H}_{0}^f \mid \text{data}) + p(\mathcal{H}_{1}^f \mid \text{data})}}_{\substack{\text{Posterior inclusion odds}\\{\text{for heterogeneity}}}}
\;\; \bigg/ \;\; 
\underbrace{ \frac{p(\mathcal{H}_{0}^r) + p(\mathcal{H}_{1}^r)}{p(\mathcal{H}_{0}^f) + p(\mathcal{H}_{1}^f)}}_{\substack{\text{Prior inclusion odds}\\{\text{for heterogeneity}}}}.
\end{equation}

The posterior BMA distribution for the effect, assuming it is present, is defined as a mixture of posterior distributions under the fixed-effect alternative hypothesis model weighted by its posterior probability and the random-effect alternative hypothesis model weighted by its posterior probability,
\begin{equation}
\label{eq:posterior_effect}
    p(\mu \mid \text{data}, \mathcal{H}_{1}) \propto p(\mu \mid \mathcal{H}_{1}^f, \text{data}) \times p(\mathcal{H}_{1}^f \mid \text{data}) + p(\mu \mid \mathcal{H}_{1}^r, \text{data}) \times p(\mathcal{H}_{1}^r \mid \text{data}).
\end{equation}
Similarly, the posterior BMA distribution for the heterogeneity, assuming it is present, is defined as a mixture of posterior distributions under the random-effect null hypothesis model weighted by its posterior probability and the random-effect alternative hypothesis model weighted by its posterior probability,
\begin{equation}
\label{eq:posterior_heterogeneity}
    p(\tau \mid \text{data}, \mathcal{H}^r) \propto p(\tau \mid \mathcal{H}_{0}^r, \text{data}) \times p(\mathcal{H}_{0}^r \mid \text{data}) + p(\tau \mid \mathcal{H}_{1}^r, \text{data}) \times p(\mathcal{H}_{1}^r \mid \text{data}).
\end{equation}

\section{Prior Distributions}

We followed the same procedure as in the previous study where we developed prior distributions for continuous outcomes\cite{bartos2021bayesian}; we downloaded data from the Cochrane Database of Systematic Reviews (CDSR), we split the data equally into a training and testing data set (50/50), we used the training set to obtain prior distributions and the test set to select the best performing prior distributions. Finally. we combined the data to propose prior distributions based on the complete data set. One exception was that in the present work we no longer consider the Cauchy prior distribution for the meta-analytic mean parameter and the Uniform prior distribution for the heterogeneity parameter as they were dominated by the remaining prior distributions.

\subsection{Data set}
As previously, we used trial data obtained from the CDSR which is the leading journal and database for systematic reviews in health care. We identified all systematic reviews in the CDSR through PubMed with the NCBI's EUtils API (query: “Cochrane Database Syst Rev”[journal] AND (“2000/01/01”[PDAT]: “2021/01/31”[PDAT]). We downloaded the XML meta‐analysis table file (rm5‐format) associated with the review’s latest version. We obtained a total of 28,579 reviews, containing 163,249 comparisons, with a grand total of 788,883 estimates. We removed estimates without the control or the treatment group and split the data according to the outcome type. This left 97,500 comparisons containing 484,128 binary trial outcomes and 1,515 comparisons with 8,149 time-to-event outcomes.

We randomly split the binary and time-to-event data into a training and testing data set. We used the highest level of grouping -- the review level -- for the 50/50 training and testing data set split to prevent information leakage and overfitting. We used the training data set to estimate the prior distributions and the testing data set to assess their predictive performance.

\subsection{Estimating prior distributions on the training data set}

\begin{figure}[h!]
\centering
{
    \resizebox{.49 \textwidth}{!}{
        \begin{tikzpicture}[node distance=3cm]
            \node (step0) [stepx] {Initial training data set: 49,247 comparisons (243,257 estimates)};
            \node (step1) [stepx, below of=step0] {6,334 comparisons (127,584 estimates)};  / 
            \node (step2) [stepx, below of=step1] {\shortstack{Final training data set: \\
                                                    log OR: 6,281 comparisons (126,794 estimates) \\
                                                    log RR: 6,215 comparisons (125,002 estimates) \\
                                                    RD: 6,318 comparisons (127,275 estimates)}};
            
            \draw [arrow] (step0) -- node[anchor=west] {Removing comparisons with fewer than 10 estimates.} (step1);
            \draw [arrow] (step1) -- node[anchor=west] {Estimating frequentist random-effects meta-analytic models.}(step2);
        \end{tikzpicture}
    }
}
\hfill 
{
    \resizebox{.49 \textwidth}{!}{
        \begin{tikzpicture}[node distance=3cm]
            \node (step0) [stepx] {Initial training data set: 731 comparisons (3,805 estimates)};
            \node (step1) [stepx, below of=step0] {100 comparisons (1,720 estimates)};
            \node (step2) [stepx, below of=step1] {Final training data set: 98 comparisons (1,692 estimates)};
            
            \draw [arrow] (step0) -- node[anchor=west] {Removing comparisons with fewer than 10 estimates.} (step1);
            \draw [arrow] (step1) -- node[anchor=west] {Estimating frequentist random-effects meta-analytic models.}(step2);
        \end{tikzpicture}
    }
}
\caption{Flowchart of the study selection procedure and data processing steps for the training data sets. Binary outcomes (left) and time to event outcomes (right).}
\label{fig:flowchart-train}
\end{figure}
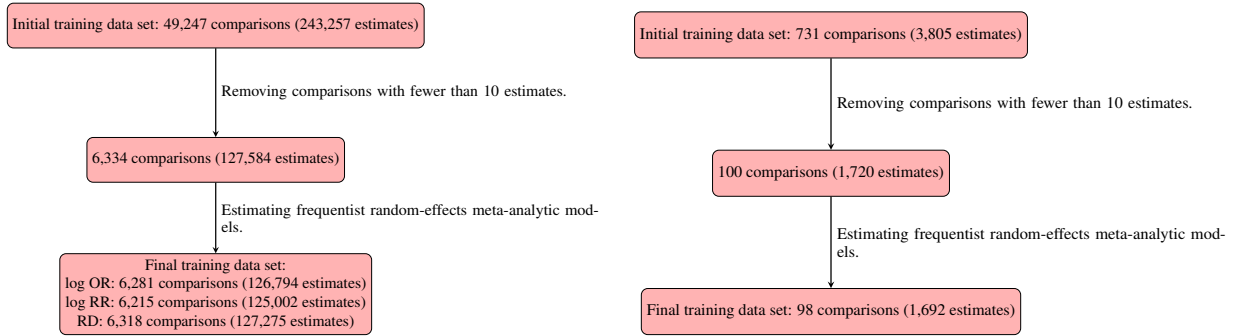

Figure~\ref{fig:flowchart-train} visualizes the data processing steps applied to the training data set. We excluded comparisons with fewer than 10 estimates to ensure that the training set yields reliable estimates of the $\mu$ and $\tau$ parameters. Then, we used the \texttt{metafor} \texttt{R} package\cite{metafor} to re-estimate all comparisons using a frequentist random-effects meta-analytic model. We used the generalized linear mixed-effects model with fixed study effects for log OR implemented in the \texttt{rma.glmm} function,\footnote{I.e., model 4 described in Jackson and colleagues\cite{jackson2018comparison} that closely resembles the Bayesian binomial-normal model in Equation~\ref{mod:binomial-normal}.} and the restricted maximum likelihood random effects model for log RR, RD, and log HR. The last row of Figure~\ref{fig:flowchart-train} summarizes the number of converged comparisons (and the corresponding number of estimates).

\begin{figure}[h]
\begin{tabular}{cc}
    \begin{minipage}{.50 \textwidth}
    \includegraphics[width=1\textwidth]{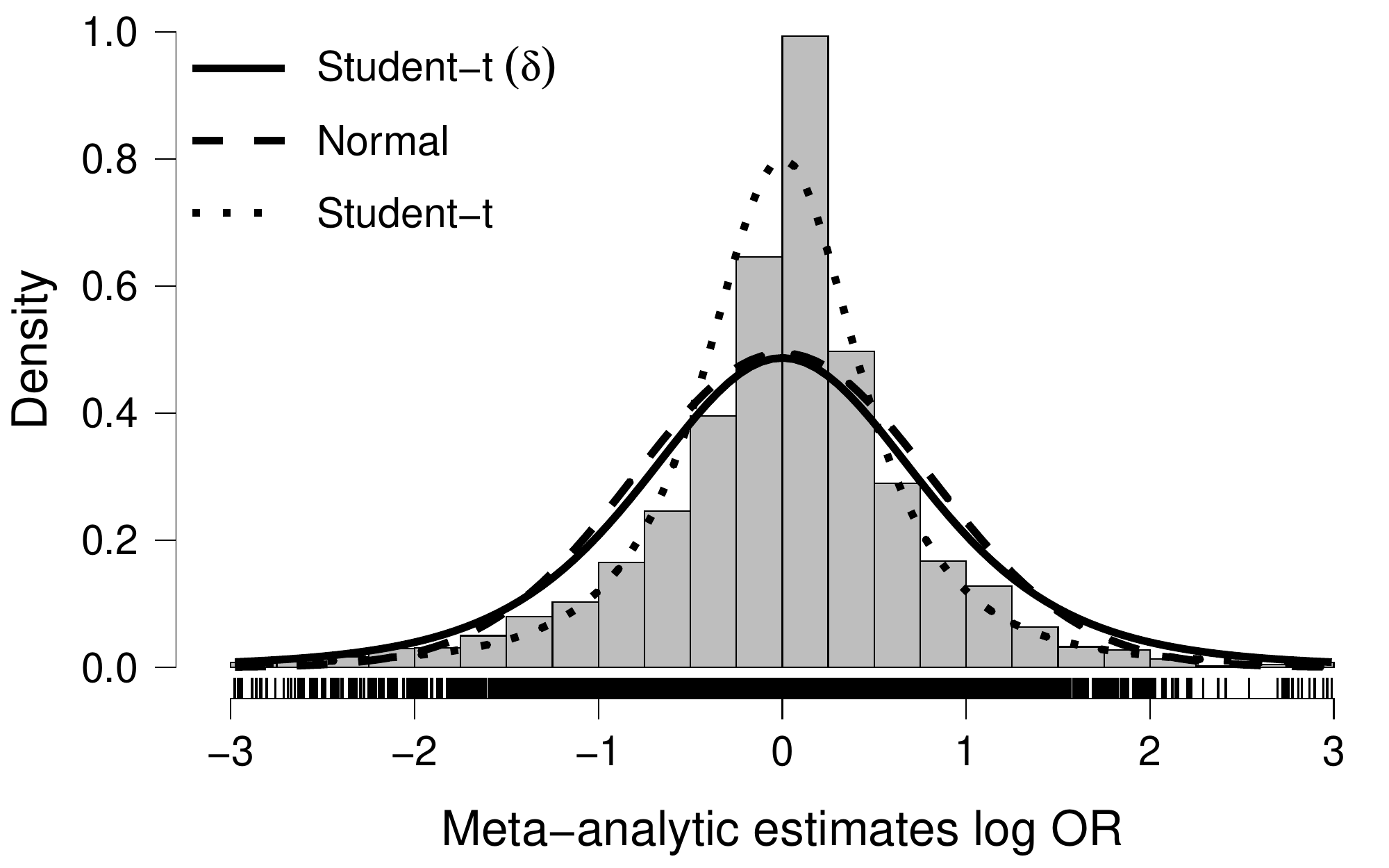}
    \end{minipage}  &
    \begin{minipage}{.50 \textwidth}
    \includegraphics[width=1\textwidth]{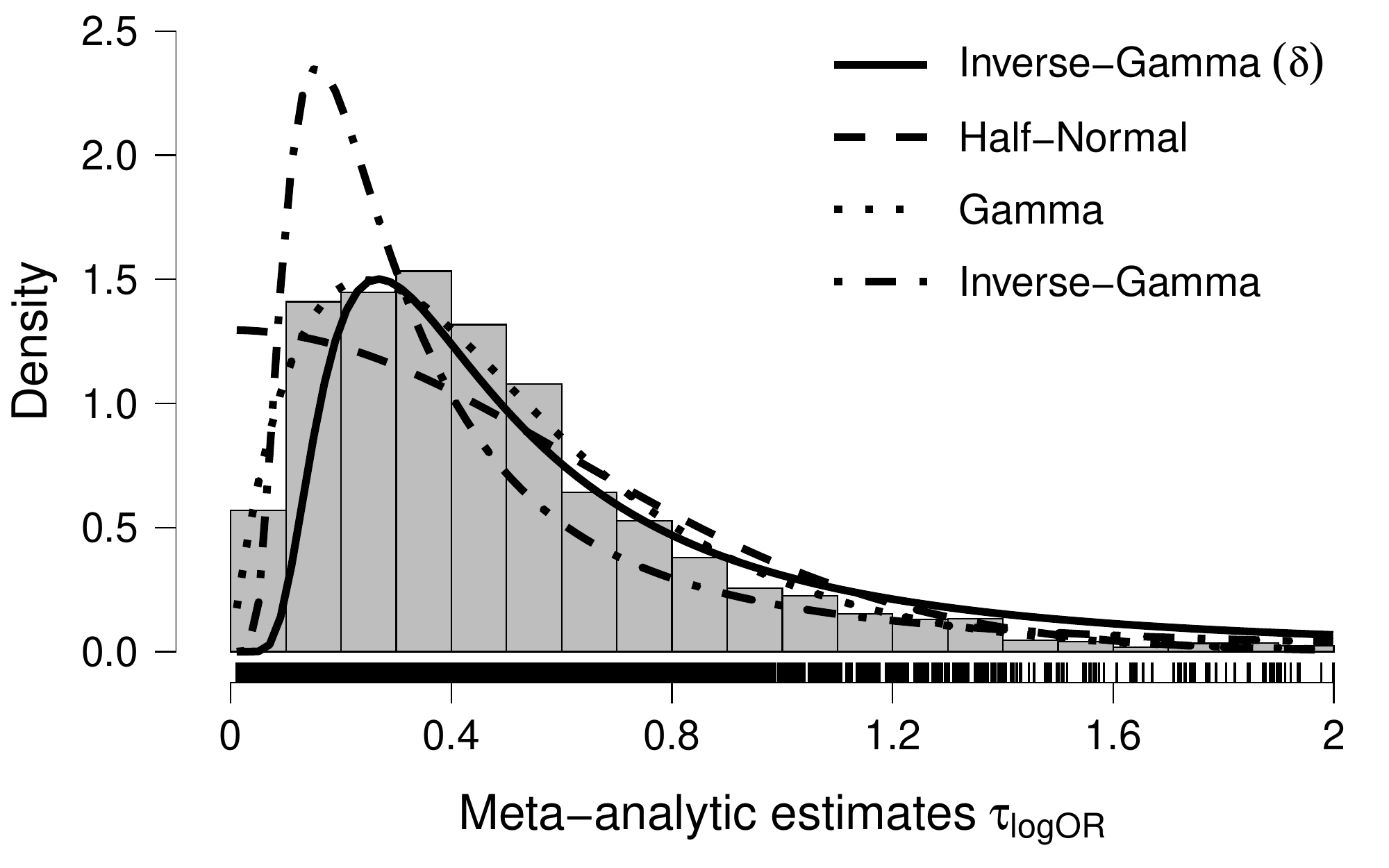}
    \end{minipage} 
    \\
    \begin{minipage}{.50 \textwidth}
    \includegraphics[width=1\textwidth]{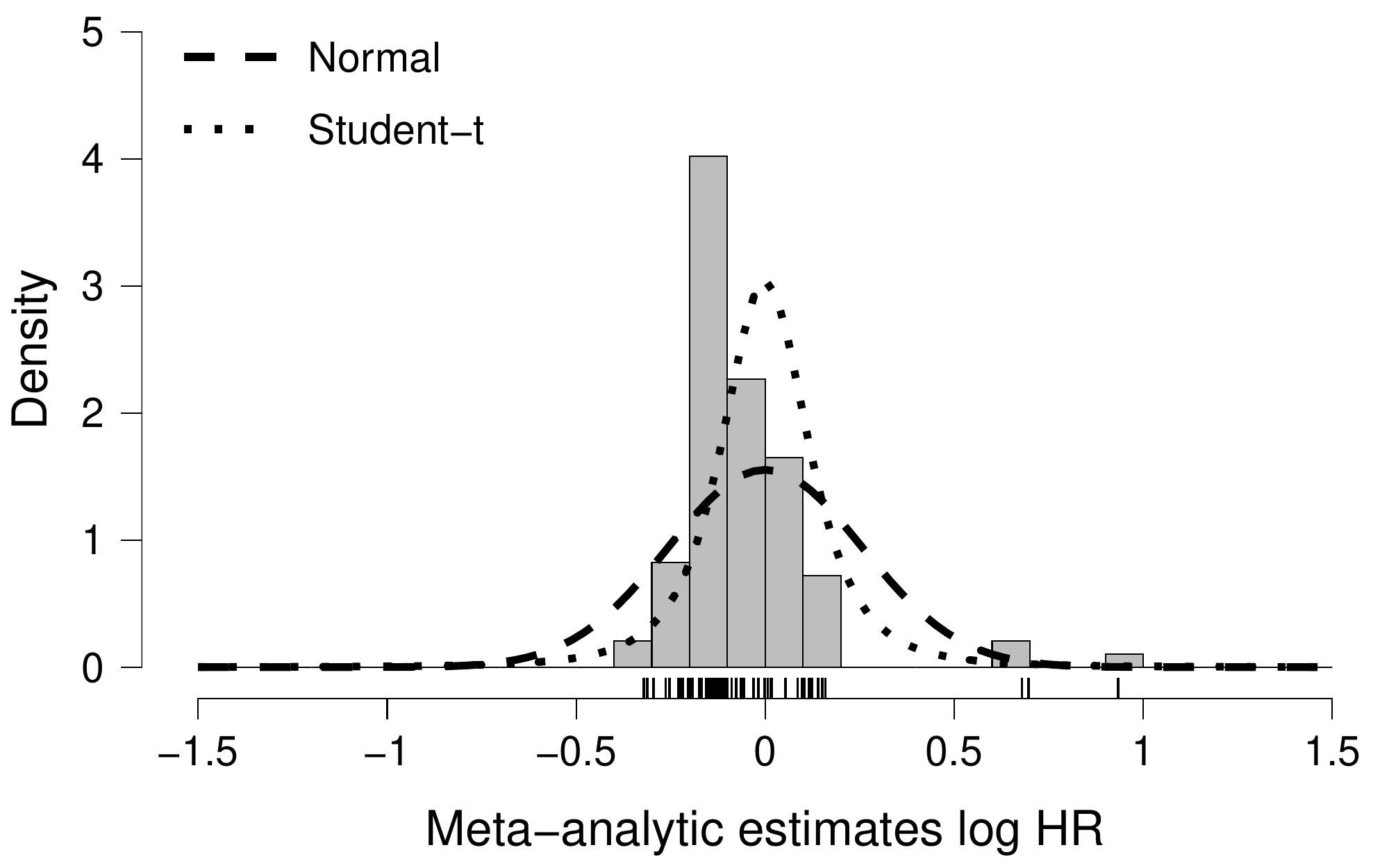}
    \end{minipage}  & 
    \begin{minipage}{.50 \textwidth}
    \includegraphics[width=1\textwidth]{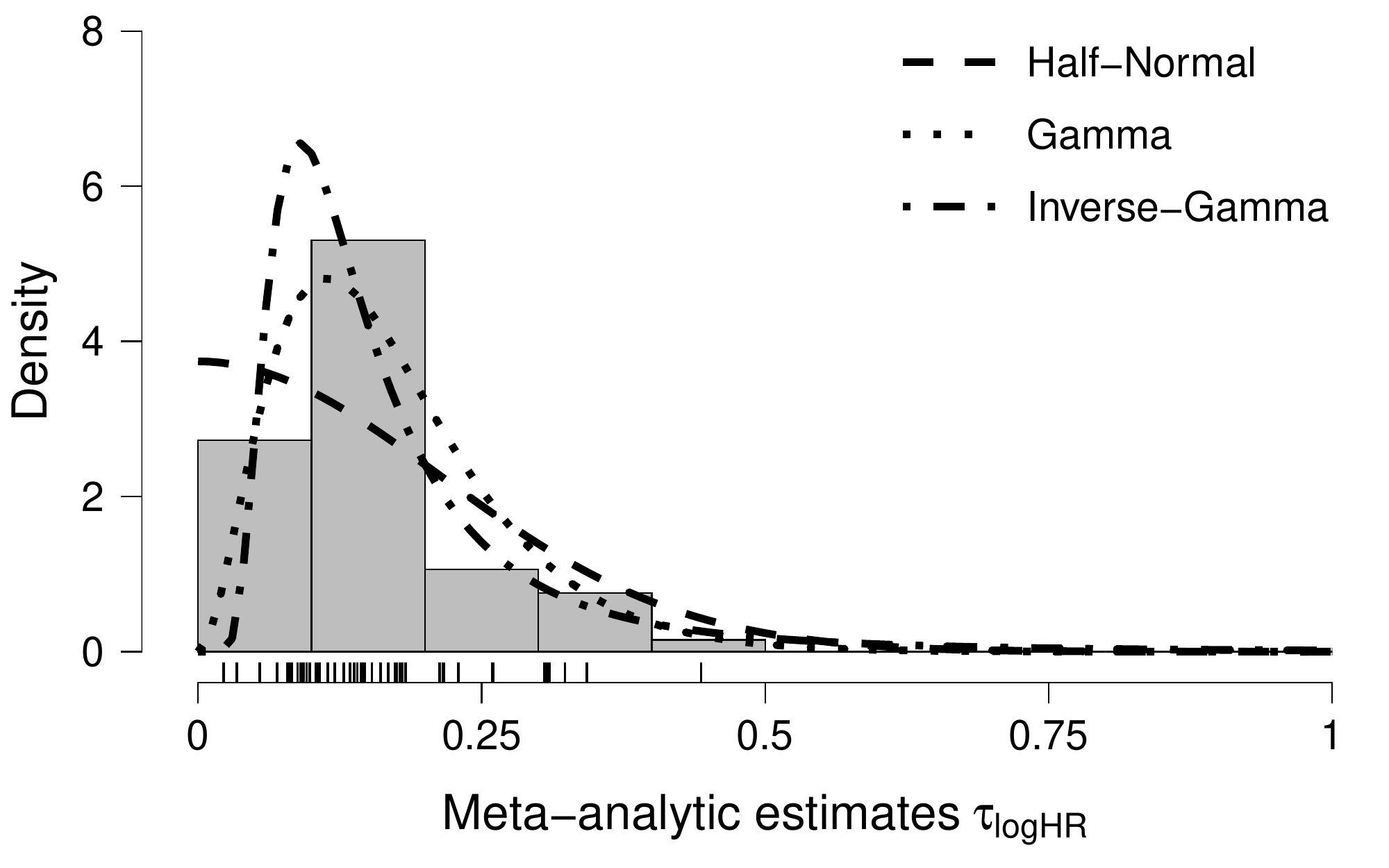}
    \end{minipage} 
    \\
\end{tabular}
\caption{Frequentist effect sizes estimates and candidate prior distributions from the training data set. Binary outcomes (log OR; first row) and time-to-event outcomes (log HR; second row). Histogram and tick marks display the estimated effect size estimates (left) and between-study standard deviation estimates (right), whereas lines represent candidate prior distributions for the population effect size parameter (left) and candidate prior distributions for the population between-study standard deviation $\tau$ (right; see Table~\ref{tab:prior-training-dataset}). 51 log OR outside of the $\pm 3$ range are not shown and 1 log HR outside of the $\pm 1.5$ range is not shown. 26 $\tau$ estimates larger than 1.5 and 2,360 $\tau$ estimates lower than 0.01 of log OR are not shown. 1 $\tau$ estimates larger than 1.5 and 31 $\tau$ estimates lower than 0.01 of log HR are not shown}
\label{fig:prior-training-dataset}
\end{figure}

We used the frequentist meta-analytic estimate obtained on the training data set to estimate prior parameter distributions for $\mu$ and $\tau$ parameters for the Bayesian meta-analytic models (Equations~\ref{mod:normal-normal} and \ref{mod:binomial-normal}; we used independent uniform distributions for the $\text{logit}(\beta_i)$ parameter as this is inconsequential for hypotheses regarding the mean effect and heterogeneity---as the prior distribution is common across the specified models).\footnote{As previously, we assumed that $\tau$ estimates lower than 0.01 are representative of $\mathcal{H}^f$. Therefore these estimates were not used to determine candidate prior distributions for $\tau$.} The first row of Figure~\ref{fig:prior-training-dataset} visualizes the histogram of mean meta-analytic estimates $\mu$ of log OR and heterogeneity $\tau$, and the second row of Figure~\ref{fig:prior-training-dataset} visualizes the histogram of mean meta-analytic estimates $\mu$ of log HR and heterogeneity $\tau$. See Figure~\ref{fig:prior-training-dataset2} in Appendix~\ref{app:other} for the corresponding visualization of log RR and RD.

We considered two distributional forms for the mean effect size parameter $\mu$: a Normal distribution centered at zero with free standard deviation parameter, and a Student's $t$-distribution centered at zero with free scale and degrees of freedom parameters. Moreover, we considered three distributions for the heterogeneity parameter $\tau$: a Half-Normal prior distribution centered at zero with free standard deviation parameter, an Inverse-Gamma distribution with free shape and scale parameters, and a Gamma distribution with free scale and shape parameters. We used the \texttt{MASS} \texttt{R} package\cite{MASS} to estimate the free parameters via maximum likelihood.

\begin{table*}[h]
  \centering
  \caption{Candidate prior distributions for the $\delta$ and $\tau$ parameters as obtained from the training set. The Student's $t$-distributions parameterized by location, scale, and degrees of freedom, the (half) normal distributions parameterized by mean and standard deviation, and the inverse-gamma and gamma distributions parameterized by shape and scale. See Figure~\ref{fig:prior-training-dataset}. The \grey{gray-colored} prior distribution corresponds to the Cohen $d$ to log OR transformed prior distribution from Bartoš and colleagues.\cite{bartos2021bayesian}}
  \label{tab:prior-training-dataset}
    \begin{tabular}{ll}
    Binary outcomes as log OR \\
    \midrule
    \grey{$\mu \sim \text{Student-t}(0, 0.78, 5)$}  & \grey{$\tau \sim \text{Inv-Gamma}(1.71, 0.73)$}  \\
    $\mu \sim \text{Normal}(0, 0.81)$               & $\tau \sim \text{Normal}_+(0, 0.62)$             \\
    $\mu \sim \text{Student-t}(0, 0.45, 2.38)$      & $\tau \sim \text{Inv-Gamma}(1.53, 0.40)$         \\
                                                    & $\tau \sim \text{Gamma}(1.99, 0.25)$             \\ 
    \bottomrule \\
    Time to event outcomes as log HR \\
    \midrule
    $\mu \sim \text{Normal}(0, 0.35)$            & $\tau \sim \text{Normal}_+(0, 0.26)$        \\
    $\mu \sim \text{Student-t}(0, 0.21, 2.57)$   & $\tau \sim \text{Inv-Gamma}(1.80, 0.21)$    \\
                                                 & $\tau \sim \text{Gamma}(1.93, 0.11)$        \\ 
    \bottomrule \\
  \end{tabular}
\end{table*}

The estimated free parameters of the prior distributions of log OR and log HR are summarized in Table~\ref{tab:prior-training-dataset}. The first prior distribution for $\mu$ and $\tau$ of log OR, in \grey{gray-colored} text, corresponds to the transformation of prior distributions obtained for Cohen's $d$ measured continuous effect sizes from Bartoš and colleagues.\cite{bartos2021bayesian} See Table~\ref{tab:prior-training-dataset2} in Appendix~\ref{app:other} for a corresponding summary of prior distributions of log RR and RD.

We observe a notable difference in the widths of the prior distributions estimated on the different effect size measures obtained from the training data set. log OR have clearly the widest prior distributions, log RR prior distributions span approximately around a half of the log OR width, while prior distributions of log HR and RD are much more narrowly concentrated around zero. This is, of course, not surprising especially in the case of RD as they are necessarily bounded to $[-1, 1]$ range.

\subsection{Assessing prior distributions on the test data set}

\begin{figure}[h!]
\centering
{
    \resizebox{.49 \textwidth}{!}{
        \begin{tikzpicture}[node distance=3cm]
            \node (step0) [stepx] {Initial training data set: 48,253 comparisons (240,871 estimates)};
            \node (step1) [stepx, below of=step0] {21,782 comparisons (206,079 estimates)};  / 
            \node (step2) [stepx, below of=step1] {\shortstack{Final training data set: \\
                                                    log OR: 21,779 comparisons (205,855 estimates) \\
                                                    log RR: 21,782 comparisons (206,079 estimates) \\
                                                    RD: 21,782 comparisons (206,079 estimates)}};
            
            \draw [arrow] (step0) -- node[anchor=west] {Removing comparisons with fewer than 3 estimates.} (step1);
            \draw [arrow] (step1) -- node[anchor=west] {Estimating Bayesian meta-analytic models.}(step2);
        \end{tikzpicture}
    }
}
\hfill 
{
    \resizebox{.49 \textwidth}{!}{
        \begin{tikzpicture}[node distance=3cm]
            \node (step0) [stepx] {Initial training data set: 784 comparisons (4,344 estimates)};
            \node (step1) [stepx, below of=step0] {491 comparisons (3,914 estimates)};
            \node (step2) [stepx, below of=step1] {Final training data set: 481 comparisons (4,363 estimates)};
            
            \draw [arrow] (step0) -- node[anchor=west] {Removing comparisons with fewer than 3 estimates.} (step1);
            \draw [arrow] (step1) -- node[anchor=west] {Estimating Bayesian meta-analytic models.}(step2);
        \end{tikzpicture}
    }
}
\caption{Flowchart of the study selection procedure and data processing steps for the test data sets. Binary outcomes (left) and time-to-event outcomes (right).}
\label{fig:flowchart-test}
\end{figure}
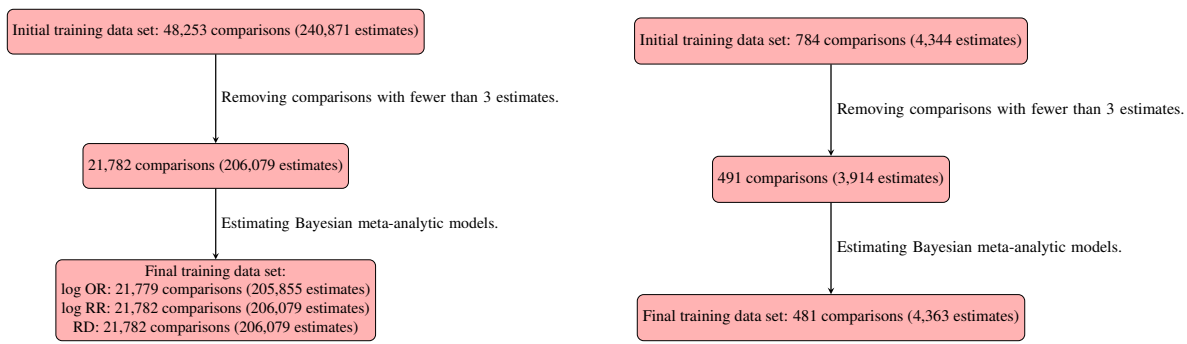

Figure~\ref{fig:flowchart-test} visualizes the data processing steps applied to the test data set. In contrast to the training data set, we included all comparisons with at least three estimates. We specified Bayesian binomial-normal meta-analytic models for log OR and Bayesian normal-normal meta-analytic models for log RR, RD, and log HR. For each effect size, we created models corresponding to all combinations of the considered prior distributions for effect size and heterogeneity (including the null hypothesis models of no effect and no heterogeneity). For instance, we estimated $4 \times 5$ meta-analytic models for log OR (a $\mu = 0$ prior distribution assuming no effect + three informed distributions for the $\mu$ parameter) $\times$ (a $\tau = 0$ prior distribution assuming no heterogeneity + four informed prior distributions for the $\tau$ parameter), as depicted in the first part of Table~\ref{tab:prior-training-dataset}.

We implemented the binomial-normal model in the \texttt{BiBMA} function and we estimated the normal-normal model via the \texttt{NoBMA} function in the \texttt{RoBMA} \texttt{R} package.\cite{RoBMA3} The \texttt{RoBMA} \texttt{R} package provides implementations for a wide range of highly modifiable Bayesian meta-analytic models estimated via MCMC using \texttt{JAGS},\cite{JAGS} computes marginal likelihoods via bridge sampling using the \texttt{bridgesampling} \texttt{R} package,\cite{bridgesampling, gronau2017tutorial} and combines the models via Bayesian model-averaging tools implemented in the \texttt{BayesTools} \texttt{R} package.\cite{BayesTools} Note that only three Bayesian binomial-normal meta-analytic models of log OR and ten normal-normal meta-analytic models of log HR did not converge.

\subsubsection{Predictive performance of the competing prior distributions}

First, we investigated the predictive performance of the competing prior distributions. We compared the specified prior distributions for each parameter separately, averaging over the specified prior distributions for the remaining parameter. For instance, when considering log OR and comparing the predictive performance of the three competing prior distributions for the $\mu$ parameter, for each specified prior distribution we averaged its performance across all four prior distributions for the $\tau$ parameter. 

\begin{table*}[h]
    \centering
    \caption{Ranking totals for each prior distribution in $\mathcal{H}_1^r$ based on the test set. The numbers indicate how many times a specific prior distribution attained a specific posterior probability rank. Rank `1' represents the best performance. The rankings reflect predictive adequacy that is model-averaged across the possible prior distribution configurations of the other parameter.}
    \label{tab:ranking-priors}
    \begin{tabular}{@{}lrrrrrr@{}}
        \multicolumn{7}{l}{Binary outcomes as log OR} \\
        \midrule
                                                  & \multicolumn{4}{c}{Rank}  &       &             \\
        Prior distribution                        &    1 &    2 &    3 &    4 & PrMP* & AV. PoMP**  \\
        \\
        \multicolumn{7}{l}{Parameter $\delta$}                                                      \\
        \midrule
        \grey{$\text{Student-t}(0, 0.78, 5)$}     &  1672 &  4761 & 15346 & --- & 0.33 & 0.32 \\ 
        $\text{Normal}(0, 0.81)$                  &  5482 & 15947 &   350 & --- & 0.33 & 0.33 \\ 
        $\text{Student-t}(0, 0.45, 2.38)$         & 14625 &  1071 &  6083 & --- & 0.33 & 0.35 \\ 
        \\
        \multicolumn{7}{l}{Parameter $\tau$} \\
        \midrule
        \grey{$\text{Inv-Gamma}(1.71, 0.73)$}   &  2011 &  1244 & 2857 & 15667 & 0.25 & 0.23 \\ 
        $\text{Normal}_+(0, 0.62)$              &  2309 & 12680 & 6072 &   718 & 0.25 & 0.26 \\ 
        $\text{Inv-Gamma}(1.53, 0.40)$          &  8915 &  2354 & 5598 &  4912 & 0.25 & 0.26 \\  
        $\text{Gamma}(1.99, 0.25)$              &  8544 &  5501 & 7252 &   482 & 0.25 & 0.26 \\  
        \bottomrule
        \\ 
        \multicolumn{7}{l}{Time to event outcomes as log HR} \\
        \midrule
                                                  & \multicolumn{4}{c}{Rank}  &       &             \\
        Prior distribution                        &    1 &    2 &    3 &    4 & PrMP* & AV. PoMP**  \\
        \\
        \multicolumn{7}{l}{Parameter $\mu$} \\
        \midrule
        $\text{Normal}(0, 0.35)$                    & 214 & 267 & ---  & ---  & 0.50 & 0.49 \\ 
        $\text{Student-t}(0, 0.21, 2.57)$           & 267 & 214 & ---  & ---  & 0.50 & 0.51 \\ 
        \\
        \multicolumn{7}{l}{Parameter $\tau$} \\
        \midrule
        $\text{Normal}_+(0, 0.26)$                & 148 & 118 & 215 & ---  & 0.33 & 0.33 \\ 
        $\text{Inv-Gamma}(1.80, 0.21)$            & 184 & 210 &  87 & ---  & 0.33 & 0.34 \\ 
        $\text{Gamma}(1.93, 0.11)$                & 149 & 153 & 179 & ---  & 0.33 & 0.33 \\ 
        \bottomrule
        
        \multicolumn{7}{l}{*Prior model probability} \\
        \multicolumn{7}{l}{**Average posterior model probability} \\
    \end{tabular}
\end{table*}

\begin{figure}[h]
\begin{tabular}{cc}
    \begin{minipage}{.50 \textwidth}
    \includegraphics[width=1\textwidth]{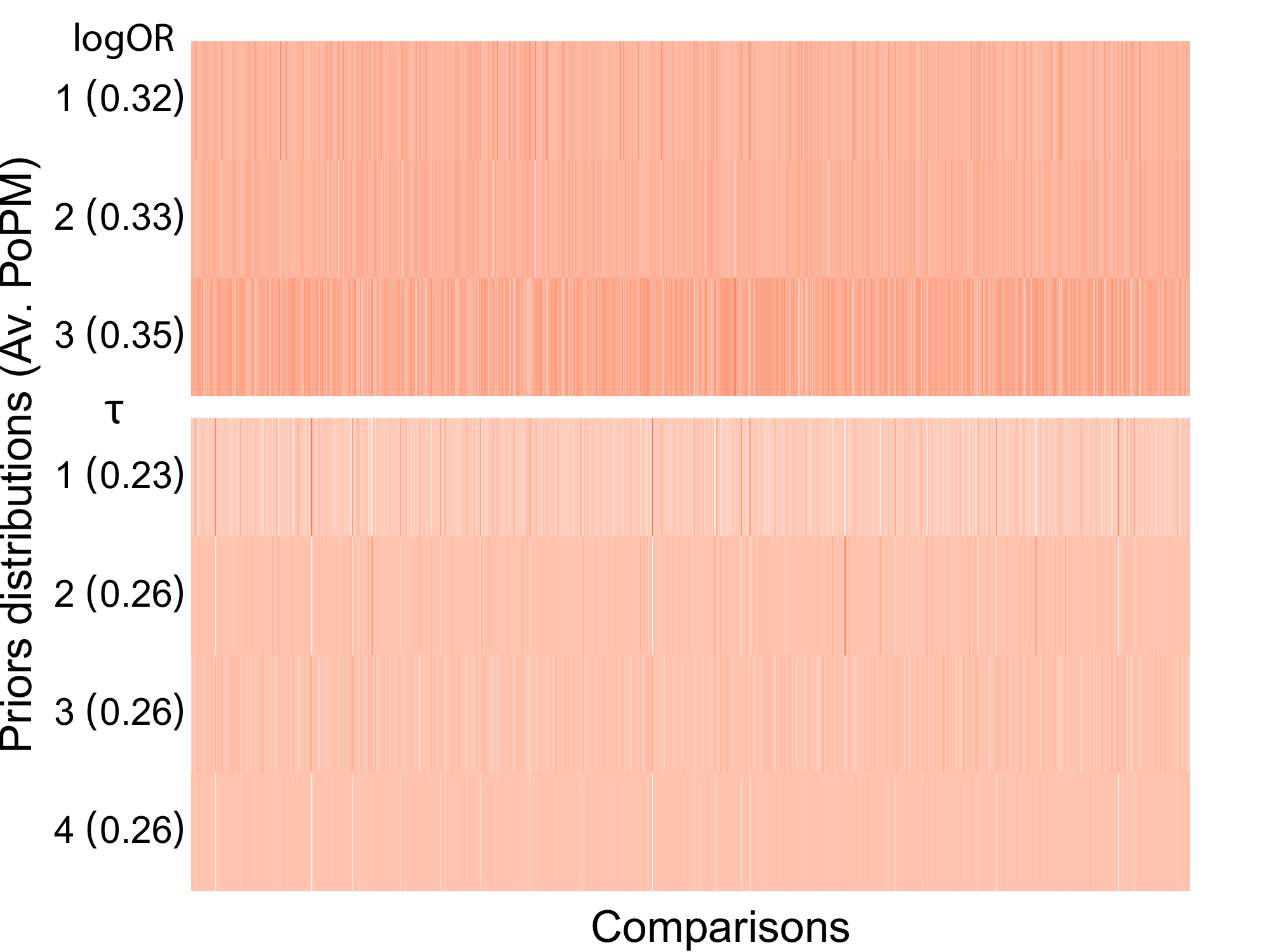}
    \end{minipage}  &
    \begin{minipage}{.50 \textwidth}
    \includegraphics[width=1\textwidth]{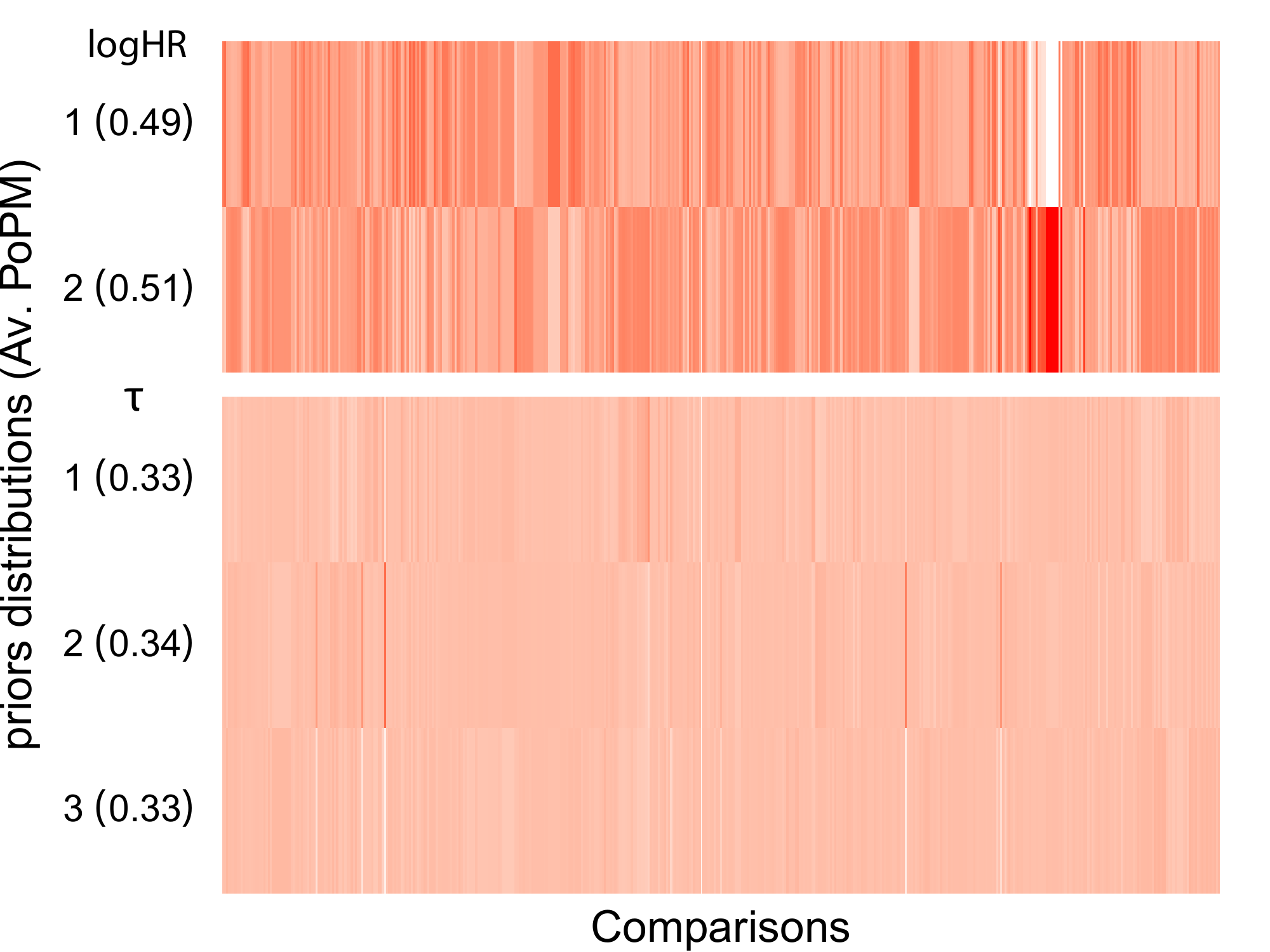}
    \end{minipage} 
\end{tabular}
\caption{Posterior model probability of the competing prior distributions for log OR and log HR. For each comparison, the color gradient ranges from white (low posterior probability) to dark red (high posterior probability). The numbers in parentheses are the averaged posterior probabilities across all comparisons.}
\label{fig:posterior-priors}
\end{figure}

Table~\ref{tab:ranking-priors} summarizes the predictive performance of the competing prior distributions for log OR and log HR in terms of ranks and the average posterior model probability, the posterior model probabilities averaged across comparisons, and Figure~\ref{fig:posterior-priors} visualizes the posterior model probabilities for each prior configuration across the comparisons for log OR and log HR. See Table~\ref{tab:ranking-priors2} and Figure~\ref{fig:posterior-priors2} in Appendix~\ref{app:other} for the corresponding summary of log RR and RD.

Across all effect size measures, we find that the more informed Student's $t$-distribution for the mean effect size parameter $\mu$ outranks the less informed normal prior distribution (most of the best ranks in Tables~\ref{tab:ranking-priors} and \ref{tab:ranking-priors2}). The dominance of the Student's t-distribution is the most notable for binary outcomes, especially RD, with a smaller difference in performance in the time-to-event outcomes. The same patern is also visible from visualization of the posterior model probabilities where the Student's $t$-distribution is painted in a sligly darker color signaling higher posterior model probabilities.

In the case of the heterogeneity parameter $\tau$ we find that the results are a bit more mixed. The more informed Inverse-Gamma and Gamma distribution attain more of the best but also the worst ranks, with an essentially indistinguishable average posterior model probability. The similar performance of Inverse-Gamma and Gamma distributions is not unexpected given their similar shapes, with the Gamma distribution slightly less peaked at small values of the heterogeneity parameter $\tau$ (Figures~\ref{fig:prior-training-dataset} and ~\ref{fig:prior-training-dataset2}). Furthermore, we find that the Half-Normal distribution outperforms the Gamma and Inverse-Gamma distributions for RD, most likely due to the shorter tail in the range-restricted effect size measure.

Finally, we find that the informed prior distributions for log OR transformed from Cohen's $d$ measured continuous outcomes slightly outperform the natively estimated informed prior distributions. This might be due to inherent differences between binary and continuous outcomes studies as well as the larger training data set presented in the current study. 

\subsubsection{Predictive performance of model types}

Second, we investigate the predictive performance of the competing model types. Since all but the fixed effect null hypothesis model $\mathcal{H}_0^f$ could use different (combinations of) prior distributions, we averaged the predictive performance across all employed prior distribution specifications. For instance, when considering the random effects alternative hypothesis $\mathcal{H}_1^r$ for log OR, we averaged the predictive performance across the twelve possible prior distribution specifications (three prior distributions for the $\mu$ parameter and four prior distributions for the $\tau$ parameter). 

\begin{table*}[h]
    \centering
    \caption{Ranking totals for each model type in the test set. The numbers indicate how many times a specific model type attained a specific posterior probability rank. Rank `1' represents the best performance. The rankings reflect predictive adequacy that is model-averaged across the possible prior distribution configurations.}
    \label{tab:ranking-models}
    \begin{tabular}{@{}lrrrrrr@{}}
    
        \multicolumn{7}{l}{Binary outcomes as log OR} \\
        \midrule
                             & \multicolumn{4}{c}{Rank}       &       &             \\
        Model                &      1 &     2 &     3 &     4 & PrMP* & AV. PoMP**  \\ 
        \midrule
        $\mathcal{H}_0^f$    &   8665 & 1704 & 1808 & 9599 & 0.25 & 0.22 \\
        $\mathcal{H}_1^f$    &   5287 & 4373 & 9715 & 2401 & 0.25 & 0.25 \\ 
        $\mathcal{H}_0^r$    &   2915 & 9576 & 8391 &  897 & 0.25 & 0.22 \\ 
        $\mathcal{H}_1^r$    &   4912 & 6126 & 1862 & 8879 & 0.25 & 0.31 \\ 
        \bottomrule
        \\
        \multicolumn{7}{l}{Time to event outcomes as log HR} \\
        \midrule
                             & \multicolumn{4}{c}{Rank}       &       &     \\
        Model                &    1 &   2 &   3 &   4 & PrMP* & AV. PoMP**  \\
        \midrule
        $\mathcal{H}_0^f$    &   101 &  36 &  51 & 293 & 0.25 & 0.13 \\ 
        $\mathcal{H}_1^f$    &   130 & 141 & 143 &  67 & 0.25 & 0.28 \\ 
        $\mathcal{H}_0^r$    &    81 & 134 & 249 &  17 & 0.25 & 0.21 \\ 
        $\mathcal{H}_1^r$    &   169 & 170 &  38 & 104 & 0.25 & 0.38 \\ 
        \bottomrule
  
        \multicolumn{7}{l}{*Prior model probability} \\
        \multicolumn{7}{l}{**Average posterior model probability} \\
    \end{tabular}
\end{table*}

\begin{figure}[h]
\begin{tabular}{cc}
    \begin{minipage}{.50 \textwidth}
    \includegraphics[width=1\textwidth]{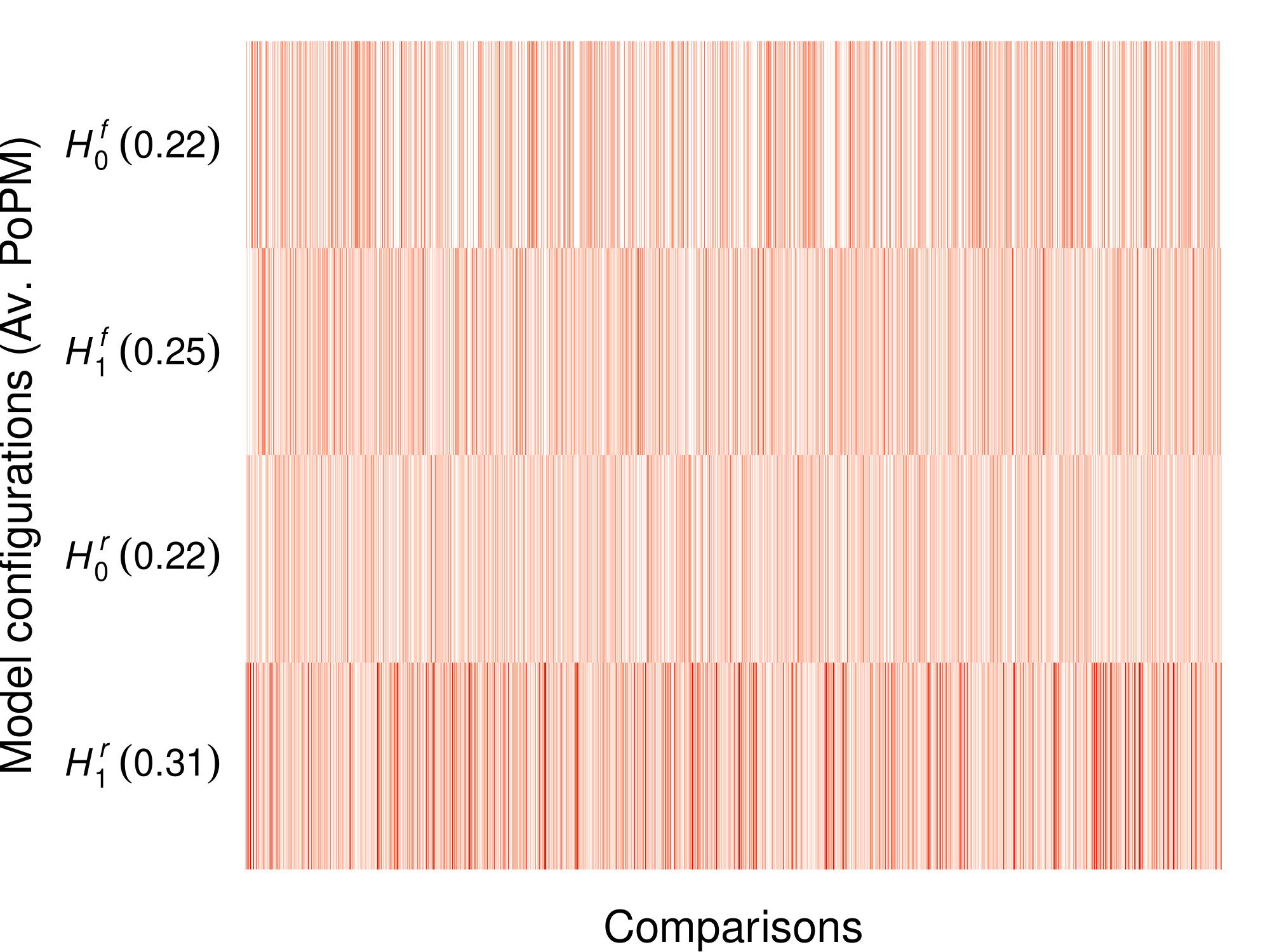}
    \end{minipage}  &
    \begin{minipage}{.50 \textwidth}
    \includegraphics[width=1\textwidth]{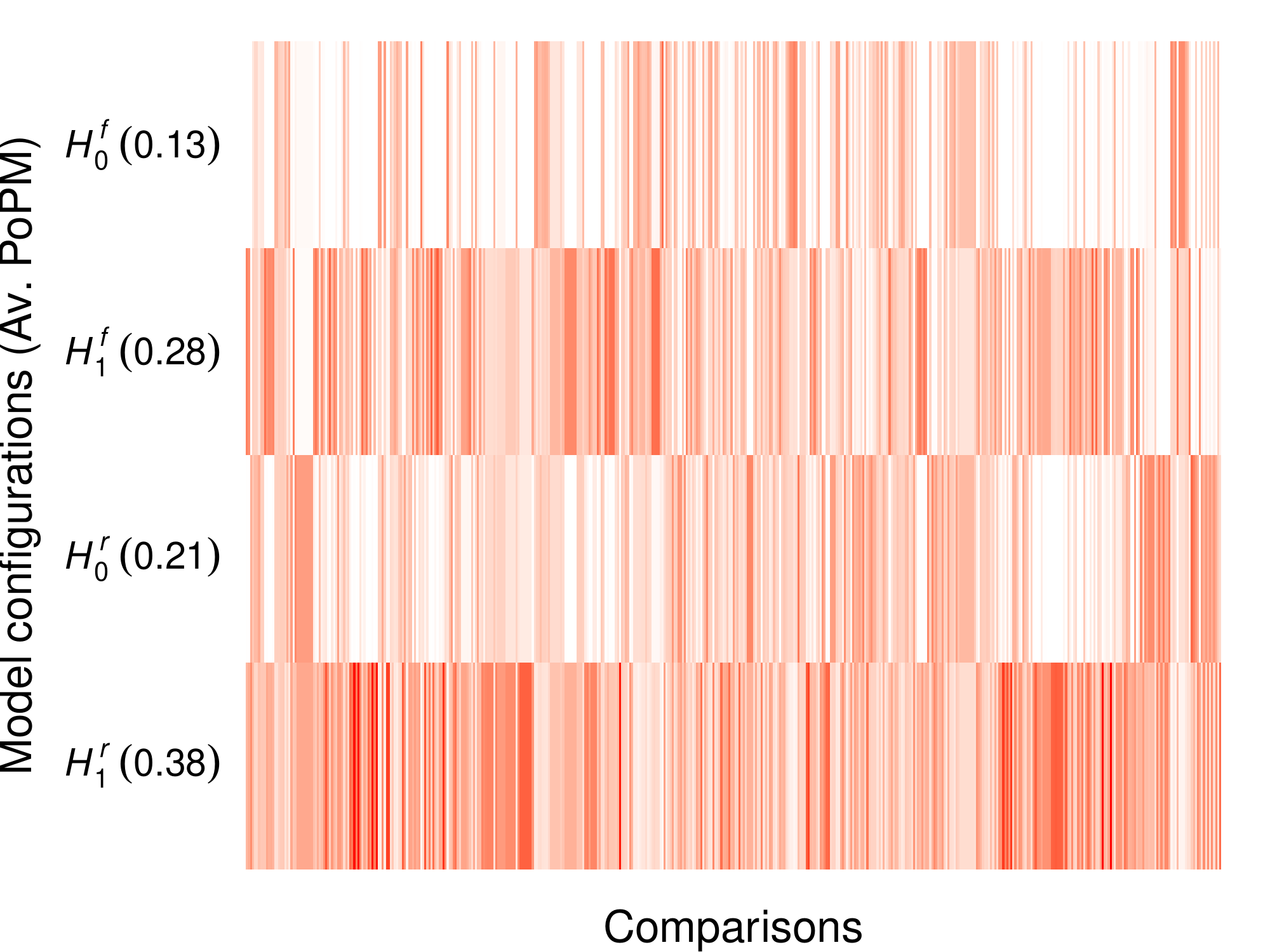}
    \end{minipage} 
\end{tabular}
\caption{Posterior model probability of the competing prior model types for log OR and log HR. For each comparison, the color gradient ranges from white (low posterior probability) to dark red (high posterior probability). The numbers in parentheses are the averaged posterior probabilities across all comparisons.}
\label{fig:posterior-hypotheses}
\end{figure}

Table~\ref{tab:ranking-models} summarizes the predictive performance of the competing model types for log OR and log HR in terms of ranks and the average posterior model probability and Figure~\ref{fig:posterior-hypotheses} visualizes the posterior model probabilities for each model type across the comparisons. Across all effect size measures, we find that the random effects alternative hypothesis $\mathcal{H}_1^r$ attains the highest posterior model probability. However, the fixed effect null hypothesis model $\mathcal{H}_0^f$ dominates in terms of the best rank in all effect sizes measures derived from binary outcomes. The generally worst performing model is the random effects null hypothesis $\mathcal{H}_0^r$ which receives the lowest posterior model probability in all effect size measures but log HR. See Table~\ref{tab:ranking-models2} and Figure~\ref{fig:posterior-hypotheses2} in Appendix~\ref{app:other} for the corresponding summary of log RR and RD.

\subsubsection{Predictive performance of hypotheses}

Third, we investigate the predictive performance of the competing hypotheses; the presence vs absence of the effect ($\mathcal{H}_1$ vs $\mathcal{H}_0$) and the random vs fixed effects models ($\mathcal{H}^r$ vs $\mathcal{H}^f$). As previously, the hypotheses were composed from multiple models and prior distributions, therefore, we averaged across the possible combinations of prior distributions. 

\begin{figure}[h]
\begin{tabular}{cc}
    \begin{minipage}{.50 \textwidth}
    \includegraphics[width=1\textwidth]{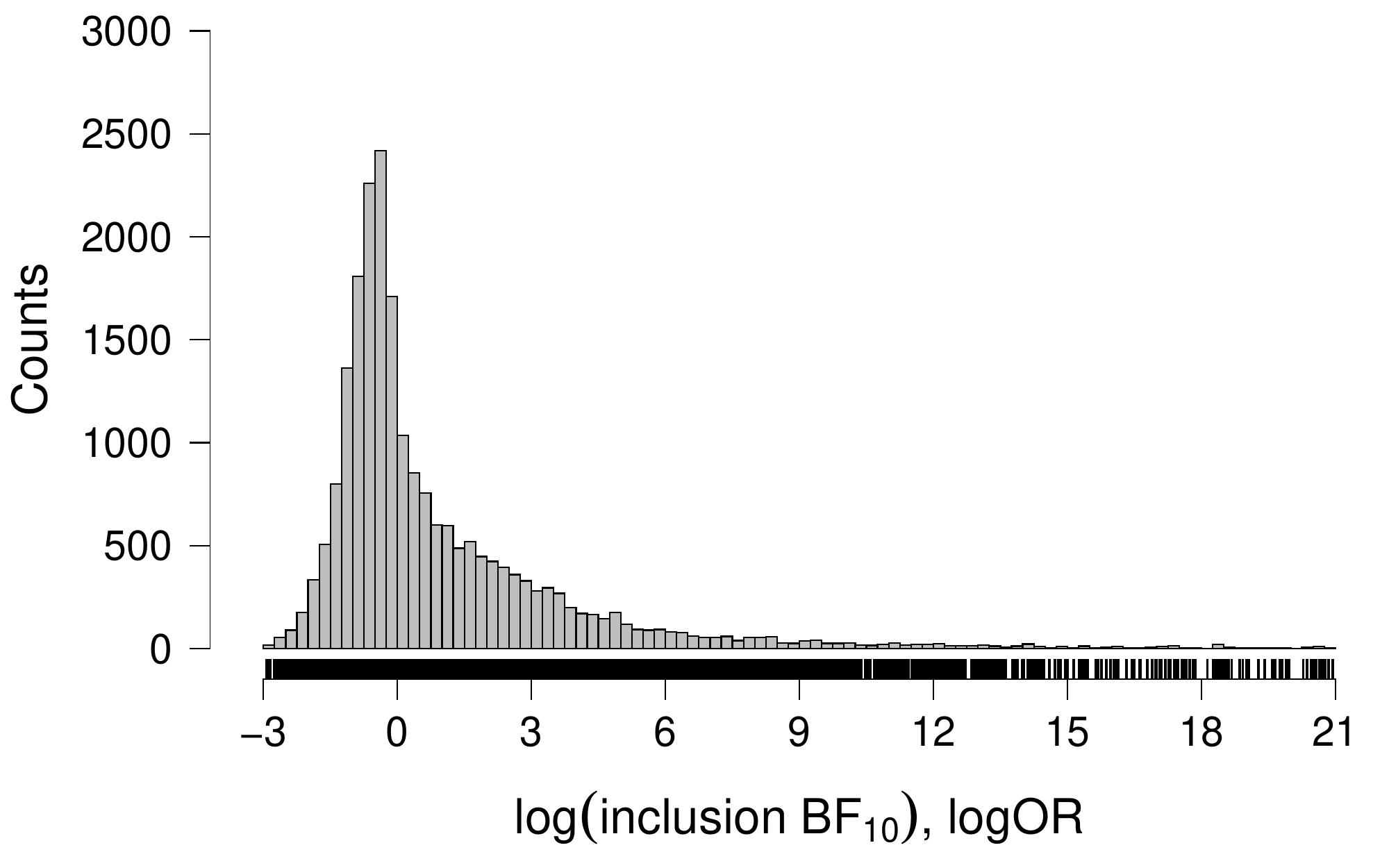}
    \end{minipage}  &
    \begin{minipage}{.50 \textwidth}
    \includegraphics[width=1\textwidth]{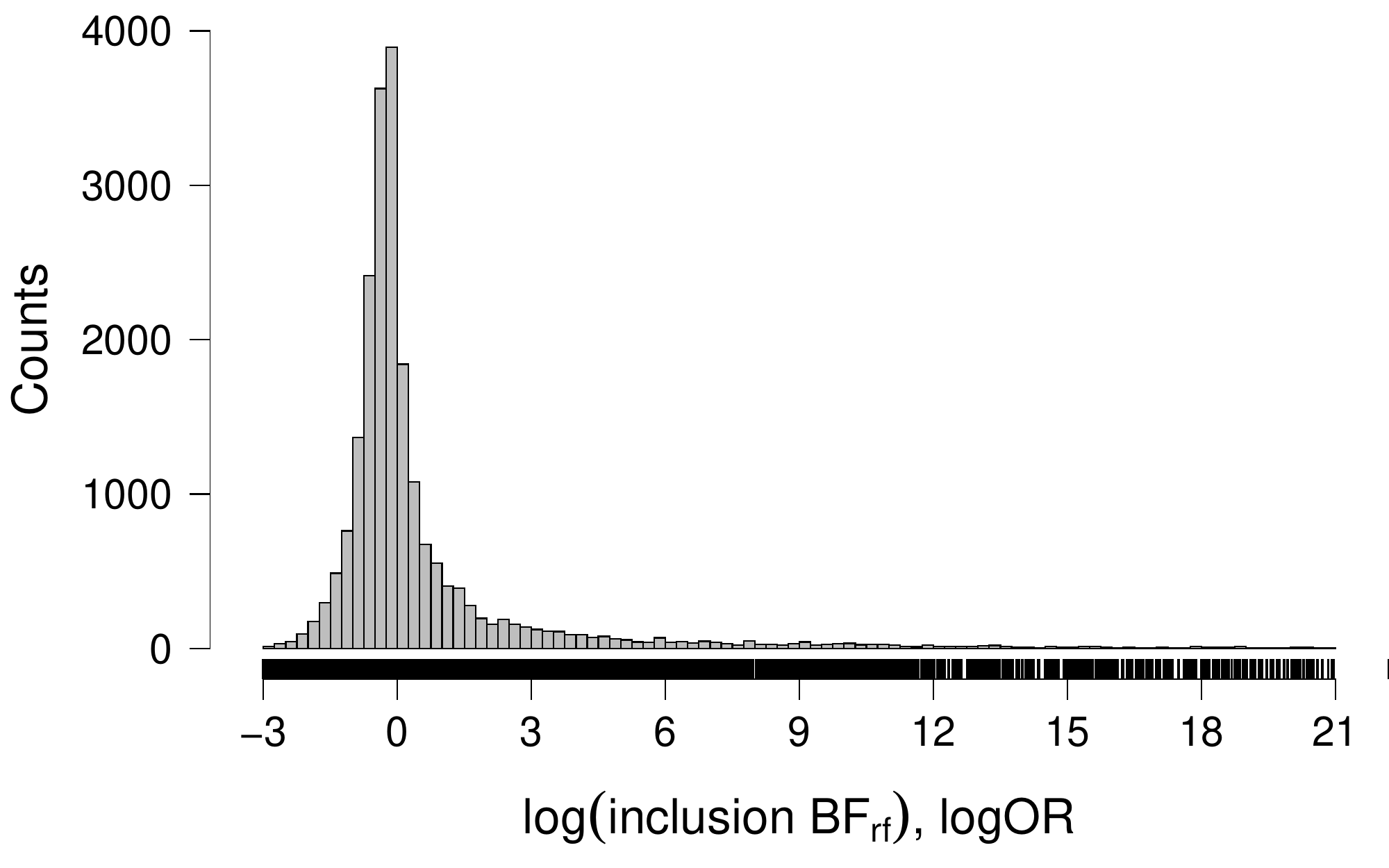}
    \end{minipage}  
    \\
    \begin{minipage}{.50 \textwidth}
    \includegraphics[width=1\textwidth]{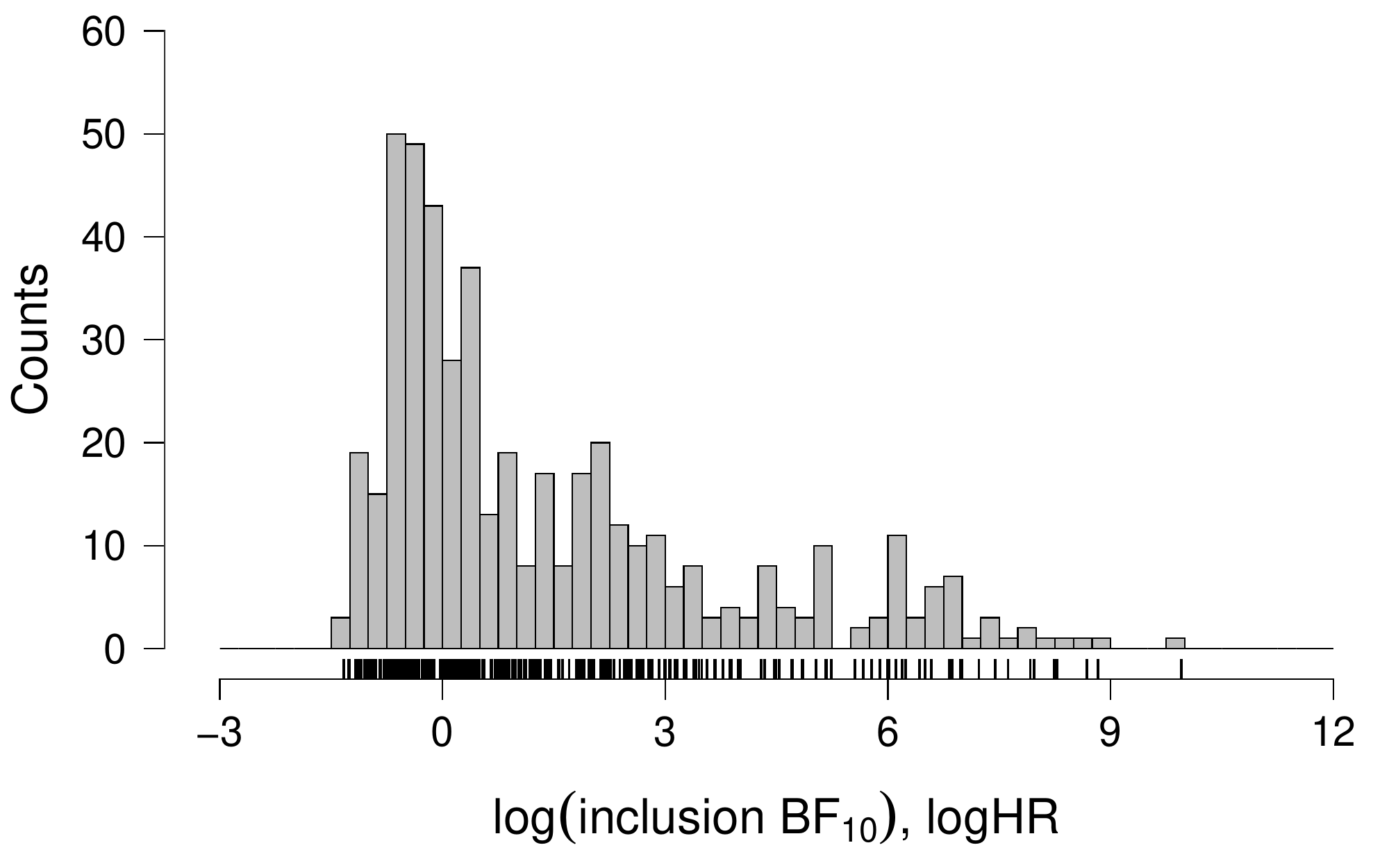}
    \end{minipage}  &
    \begin{minipage}{.50 \textwidth}
    \includegraphics[width=1\textwidth]{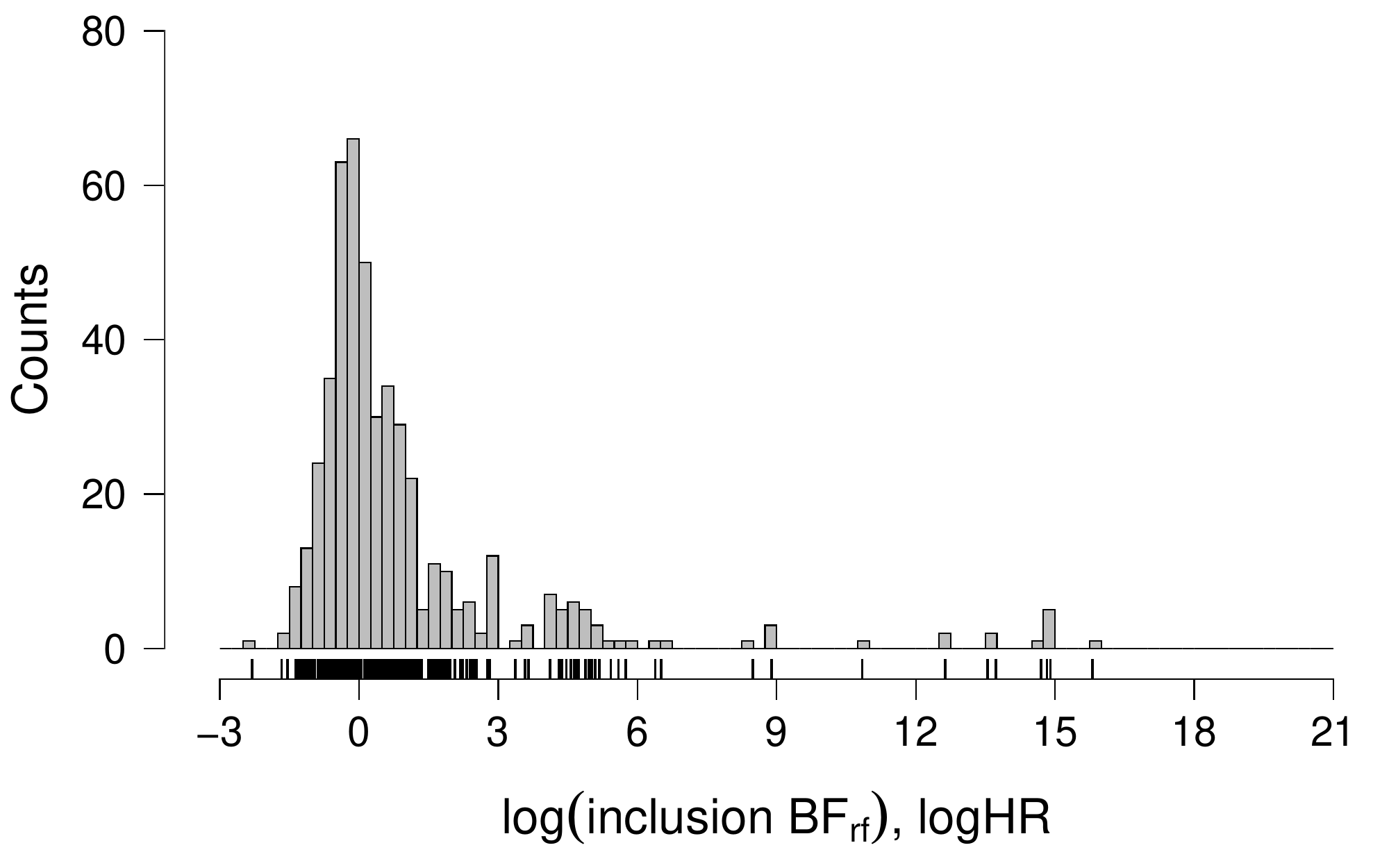}
    \end{minipage}  
\end{tabular}
\caption{Inclusion Bayes factors in favor of the presence of a treatment effect (left) and in favor of the presence of across-study heterogeneity (right) in the test set. Binary outcomes (log OR; first row) and time-to-event outcomes (log HR; second row). In log OR; for the presence of the effect, 121 log Bayes factors larger than 21 and 12 log Bayes factors lower than -3, and for the presence of heterogeneity, 478 log Bayes factors larger than 21 and 34 log Bayes factors lower than $-3$ are not shown. In log HR; for the presence of the effect, 9 log Bayes factors larger than 12, and for the presence of heterogeneity, 2 log Bayes factors larger than 21 are not shown.}
\label{fig:inclusion-BF}
\end{figure}

Figure~\ref{fig:inclusion-BF} visualizes the distribution of inclusion Bayes factors for the effect and heterogeneity for log OR and log HR. We find the usual skewed distribution of Bayes factors, with the skew favoring inclusion of Bayes factors for the presence of the effect and heterogeneity (as it is more difficult to obtain evidence for absence). Nonetheless, in all effect size measures of binary outcomes, we find that only the minority of comparisons yields with the evidence in favor of the effect ($\text{BF}_{10} > 1$: $47.0$\% for log OR, $45.1$\% for log RR, and $40.1$\% for RD) or heterogeneity ($\text{BF}_{rf} > 1$: $39.2$\% for log OR, $34.6$\% for log RR, and $36.3$\% for RD). For log HR, the reverse is true as $62.8$\% of comparisons yield evidence in favor of the effect and $55.9$\% of comparisons yield evidence for heterogeneity. See Figure~\ref{fig:inclusion-BF2} in Appendix~\ref{app:other} for the corresponding visualization for log RR and RD.
 
\subsection{Estimating prior distributions on the complete data}

Finally, we used the complete data set to estimate empirical prior distributions for the field in general and for the specific sub-disciplines. We applied the same data processing steps as on the training data set, resulting in 12,079 comparisons of binary outcomes (255,048 estimates) and 234 comparisons of time-to-event outcomes (3,831 estimates). After estimating the frequentist meta-analytic models, we ended up with 11,964 log OR comparisons (253,054 estimates), 11,862 log RR comparisons (250,331 estimates), 12,042 RD comparisons (254,326 estimates), and 225 log HR comparisons (3,707 estimates), which we used for constructing the data-driven general and subfield-specific prior distributions.

For comparisons of binary outcomes, we estimate the general and subfield-specific empirical prior distributions jointly using Bayesian hierarchical estimation with weakly informative priors on the hyperparameters. Bayesian hierarchical modeling allows us to shrink the estimated subfield parameter values towards the grand mean, which is especially useful for subfields with relatively little information and extreme values.\cite{lee2005bayesian, mcelreath2020statistical, rouder2005introduction} We implement the hierarchical models using the \texttt{rstan} \texttt{R} package\cite{rstan} that interfaces with the Stan probabilistic modeling language.\cite{stan} We specify the models as such that all field-specific parameters of the Student's $t$-distributions (i.e., $\sigma_i$ and $\nu_i$) and the Inverse-Gamma distributions (i.e., $\alpha_i$ and $\beta_i$) are shrunk via a positive-only normal distributions. For the populational parameters, we use positive-only Cauchy distributions with scale one for the location of the Student's $t$-distribution ($\sigma \sim \text{Cauchy}_+(0, 1)$), scale ten for the degrees of freedom of the Student's $t$-distribution ($\nu \sim \text{Cauchy}_+(0, 10)$), and with scale one for the shape and scale of the Inverse-Gamma distribution ($\alpha, \beta \sim \text{Cauchy}_+(0, 1)$).

Due to the insufficient number of time-to-event comparisons to estimate prior distributions for the subfields (i.e., 225 comparisons), we provide only a general empirical prior distribution for log HR estimated via maximum likelihood.

We estimate the more informed, and generally dominant, Student's $t$-distribution for the mean effect size parameter $\mu$ of all effect size measures. For the heterogeneity parameter $\tau$ of log OR, log RR, and log HR we estimate an Inverse-Gamma distribution (the performance is generally tied with the Gamma distribution) and for RD we estimate the more dominant Half-Normal prior distribution.

\begin{table}[h]
\centering
\caption{General and subfield-specific prior distributions for log OR of individual topics from the Cochrane database of systematic reviews estimated by hierarchical regression based on the complete data set. The Student's $t$-distributions parameterized by location, scale, and degrees of freedom and the Inverse-Gamma distribution parameterized by shape and scale.}
\label{tab:priors-log-OR}
\small
\begin{tabular}{lrrrr}
Topic & Comparisons $\mu$ & Comparisons $\tau$ & Prior $\mu$ & Prior $\tau$ \\
\toprule
  Acute Respiratory Infections & 308 (5860) & 197 (4061) & Student-t(0, 0.48, 3) & Inv-Gamma(1.67, 0.45) \\ 
  Airways & 450 (8702) & 207 (3839) & Student-t(0, 0.37, 2) & Inv-Gamma(1.35, 0.27) \\ 
  Anaesthesia & 408 (9409) & 261 (6781) & Student-t(0, 0.79, 6) & Inv-Gamma(2.12, 0.86) \\ 
  Back and Neck & 14 (295) & 10 (238) & Student-t(0, 0.62, 4) & Inv-Gamma(1.84, 0.68) \\ 
  Bone, Joint and Muscle Trauma & 202 (3732) & 103 (1937) & Student-t(0, 0.44, 2) & Inv-Gamma(1.01, 0.36) \\ 
  Breast Cancer & 137 (2646) & 113 (2296) & Student-t(0, 0.39, 3) & Inv-Gamma(1.59, 0.48) \\ 
  Childhood Cancer & 1 (13) & --- & Student-t(0, 0.47, 4) & --- \\ 
  Colorectal & 230 (4120) & 141 (2685) & Student-t(0, 0.65, 4) & Inv-Gamma(1.71, 0.60) \\ 
  Common Mental Disorders & 649 (13322) & 422 (9731) & Student-t(0, 0.54, 4) & Inv-Gamma(1.63, 0.45) \\ 
  Consumers and Communication & 44 (918) & 41 (875) & Student-t(0, 0.48, 4) & Inv-Gamma(2.37, 0.86) \\ 
  Cystic Fibrosis and Genetic Disorders & 81 (1843) & 40 (964) & Student-t(0, 0.40, 3) & Inv-Gamma(1.82, 0.58) \\ 
  Dementia and Cognitive Improvement & 127 (2097) & 78 (1278) & Student-t(0, 0.49, 4) & Inv-Gamma(1.28, 0.25) \\ 
  Developmental, Psychosocial and Learning Problems & 106 (1457) & 79 (1083) & Student-t(0, 0.83, 5) & Inv-Gamma(1.83, 0.82) \\ 
  Drugs and Alcohol & 103 (2027) & 73 (1545) & Student-t(0, 0.44, 4) & Inv-Gamma(1.59, 0.42) \\ 
  Effective Practice and Organisation of Care & 59 (1106) & 44 (878) & Student-t(0, 0.51, 4) & Inv-Gamma(1.90, 0.68) \\ 
  Emergency and Critical Care & 233 (4803) & 152 (3466) & Student-t(0, 0.35, 3) & Inv-Gamma(1.46, 0.34) \\ 
  ENT & 17 (243) & 10 (136) & Student-t(0, 0.81, 4) & Inv-Gamma(1.73, 0.71) \\ 
  Epilepsy & 114 (2052) & 39 (839) & Student-t(0, 0.88, 6) & Inv-Gamma(1.71, 0.43) \\ 
  Eyes and Vision & 57 (1007) & 41 (815) & Student-t(0, 0.77, 5) & Inv-Gamma(2.09, 0.94) \\ 
  Fertility Regulation & 62 (1072) & 46 (818) & Student-t(0, 0.46, 5) & Inv-Gamma(2.21, 0.71) \\ 
  Gut & 314 (6056) & 215 (4595) & Student-t(0, 0.63, 5) & Inv-Gamma(1.94, 0.62) \\ 
  Gynaecological, Neuro-oncology and Orphan Cancer & 245 (5811) & 159 (4123) & Student-t(0, 0.53, 4) & Inv-Gamma(1.80, 0.56) \\ 
  Gynaecology and Fertility & 364 (6896) & 185 (3714) & Student-t(0, 0.40, 2) & Inv-Gamma(1.24, 0.28) \\ 
  Haematology & 155 (4546) & 101 (3080) & Student-t(0, 0.57, 4) & Inv-Gamma(2.91, 0.66) \\ 
  Heart & 979 (20760) & 579 (14085) & Student-t(0, 0.20, 2) & Inv-Gamma(1.64, 0.29) \\ 
  Heart; Vascular & 8 (178) & 7 (166) & Student-t(0, 0.95, 4) & Inv-Gamma(1.64, 0.83) \\ 
  Hepato-Biliary & 1042 (36665) & 706 (25668) & Student-t(0, 0.43, 3) & Inv-Gamma(1.58, 0.40) \\ 
  HIV/AIDS & 32 (475) & 22 (293) & Student-t(0, 0.32, 4) & Inv-Gamma(1.76, 0.36) \\ 
  Hypertension & 66 (1127) & 28 (453) & Student-t(0, 0.28, 5) & Inv-Gamma(1.25, 0.10) \\ 
  Incontinence & 78 (1479) & 53 (1023) & Student-t(0, 0.75, 3) & Inv-Gamma(2.07, 1.09) \\ 
  Infectious Diseases & 361 (6619) & 251 (4787) & Student-t(0, 0.66, 3) & Inv-Gamma(2.08, 0.86) \\ 
  Injuries & 214 (5662) & 143 (4182) & Student-t(0, 0.60, 4) & Inv-Gamma(1.52, 0.49) \\ 
  Kidney and Transplant & 479 (8828) & 274 (5226) & Student-t(0, 0.53, 4) & Inv-Gamma(1.68, 0.44) \\ 
  Lung Cancer & 76 (1416) & 68 (1294) & Student-t(0, 0.61, 5) & Inv-Gamma(2.04, 0.68) \\ 
  Metabolic and Endocrine Disorders & 130 (3092) & 78 (2190) & Student-t(0, 0.29, 2) & Inv-Gamma(0.92, 0.11) \\ 
  Methodology & 74 (2098) & 69 (1997) & Student-t(0, 0.60, 5) & Inv-Gamma(2.04, 0.50) \\ 
  Movement Disorders & 59 (1058) & 35 (708) & Student-t(0, 0.73, 5) & Inv-Gamma(2.14, 0.64) \\ 
  Multiple Sclerosis and Rare Diseases of the CNS & 29 (564) & 23 (487) & Student-t(0, 0.76, 4) & Inv-Gamma(2.09, 0.71) \\ 
  Musculoskeletal & 139 (2403) & 92 (1686) & Student-t(0, 0.59, 4) & Inv-Gamma(1.76, 0.59) \\ 
  Neonatal & 337 (6327) & 153 (2738) & Student-t(0, 0.29, 3) & Inv-Gamma(1.80, 0.42) \\ 
  Neuromuscular & 43 (739) & 20 (331) & Student-t(0, 0.70, 5) & Inv-Gamma(1.74, 0.49) \\ 
  Oral Health & 63 (990) & 45 (724) & Student-t(0, 1.13, 4) & Inv-Gamma(1.85, 0.70) \\ 
  Pain, Palliative and Supportive Care & 315 (6062) & 204 (4292) & Student-t(0, 1.00, 7) & Inv-Gamma(1.50, 0.40) \\ 
  Pregnancy and Childbirth & 1307 (24397) & 832 (16588) & Student-t(0, 0.38, 2) & Inv-Gamma(1.73, 0.47) \\ 
  Schizophrenia & 609 (13515) & 405 (9919) & Student-t(0, 0.58, 4) & Inv-Gamma(1.92, 0.69) \\ 
  Sexually Transmitted Infections & 25 (318) & 19 (236) & Student-t(0, 0.61, 2) & Inv-Gamma(1.72, 0.47) \\ 
  Skin & 142 (2309) & 76 (1294) & Student-t(0, 0.81, 2) & Inv-Gamma(1.64, 0.49) \\ 
  Stroke & 299 (5310) & 131 (2357) & Student-t(0, 0.22, 2) & Inv-Gamma(1.36, 0.21) \\ 
  Tobacco Addiction & 213 (4990) & 168 (4079) & Student-t(0, 0.49, 5) & Inv-Gamma(2.51, 0.63) \\ 
  Urology & 89 (1673) & 60 (1271) & Student-t(0, 0.82, 5) & Inv-Gamma(1.72, 0.50) \\ 
  Vascular & 187 (2590) & 116 (1711) & Student-t(0, 0.68, 5) & Inv-Gamma(1.70, 0.45) \\ 
  Work & 14 (208) & 11 (178) & Student-t(0, 0.59, 4) & Inv-Gamma(1.79, 0.73) \\ 
  Wounds & 75 (1169) & 53 (888) & Student-t(0, 0.60, 4) & Inv-Gamma(2.16, 0.86) \\ 
  \midrule
  Pooled Estimate & 11964 (253054) & 7478 (170628) & Student-t(0, 0.58, 4) & Inv-Gamma(1.77, 0.55) \\ 
\bottomrule
\end{tabular}
\end{table}

The resulting empirical prior distributions are summarized in Table~\ref{tab:priors-log-OR} for log OR, and in Table~\ref{tab:priors-log-RR} and Table~\ref{tab:priors-RD} in Appendix~\ref{app:other} for the log RR and RD. The hierarchically pooled empirical prior distributions based on the complete data sets are $\mu \sim \text{Student-t}(0, 0.58, 4)$ and $\tau \sim \text{Inv-Gamma}(1.77, 0.55)$ for log OR, $\mu \sim \text{Student-t}(0, 0.32, 3)$ and $\tau \sim \text{Inv-Gamma}(1.51, 0.23)$ for log RR, and $\mu \sim \text{Student-t}(0, 0.03, 1)$ and $\tau \sim \text{Normal}_+(0, 0.10)$ for RD. The maximum likelihood empirical prior distributions on the complete data sets are $\mu \sim \text{Student-t}(0, 0.13, 2)$ and $\tau \sim \text{Inv-Gamma}(2.42, 0.30)$ for log HR.

We see a notable heterogeneity across topics. For instance, the prior distribution for the meta-analytic mean $\mu$ in the ``Oral Health'' topic is as much as six times wider that the prior distribution for the  ``Heart'' topic. These between-topic differences highlight the possibility for incorporating domain specific information into the statistical inference.

\section{Example: Adverse effects of honey in treating acute cough in children}

We illustrate the methodology with an example from the field of acute respiratory infections. Oduwole et al.\cite{oduwole2018honey} examined the effect of honey on treating acute cought in children. While the main analyses focused on reduction in Likert-scaled measured symptomatic relief of cough, finding possible benefits of giving honey to children, we focus on the avdverse events analysis (because it features logOR on zero cells). Specificaly, we re-examine the comparison honey and no treatment on the presence of nervousness, insomnia, hyperactivity. Oduwole et al.\cite{oduwole2018honey} found two elligible studies---with 5/35 and 2/40 events in the honey condition, and 0/39 and 0/40 events in the no treatment conditions---and reported meta-analytic effect size estimate OR = $9.40$, 95\% CI [$1.16$, $76.20$], $z = 2.10$, $p = 0.04$ of honey on the presennce of nervousness, insomnia, hyperactivity (``Analysis 3.5. Comparison 3 Adverse events, Outcome 5 Honey versus no treatment'', p. 73).

We conducted a re-examination of the Oduwole et al. meta-analysis \cite{oduwole2018honey} using the binomial-normal model, which we implemented in the open-source statistical software package JASP (jasp-stats.org) \cite{JASP17, ly2021bayesian, vandoorn2020jasp}. For the same analysis in \texttt{R} \cite{R}, we utilized the \texttt{RoBMA} package and have included the details in Appendix~\ref{app:example_R}.

To perform a BMA meta-analysis using JASP, we loaded the data and activated the ``Meta-Analysis'' module by clicking on the blue "+" button in the top right corner. Then, we selected "Meta-Analysis" from the ribbon at the top, followed by choosing "Bayesian Meta-Analysis (Binomial)" from the drop-down menu. In the left input panel, we moved the observed number of events and the number of participants in each group into the corresponding boxes and adjusted the prior distributions under the ``Prior'' tab to match the ``Acute Respiratory Infections'' subfield-specific prior distributions given in Table~\ref{tab:priors-log-OR} ($\delta \sim \mathcal{T}(0, 0.48, 3)$ and $\tau \sim \text{Inv-Gamma}(1.67, 0.45)$).

\begin{figure*}[h]
\centering
    \includegraphics[width=1\textwidth]{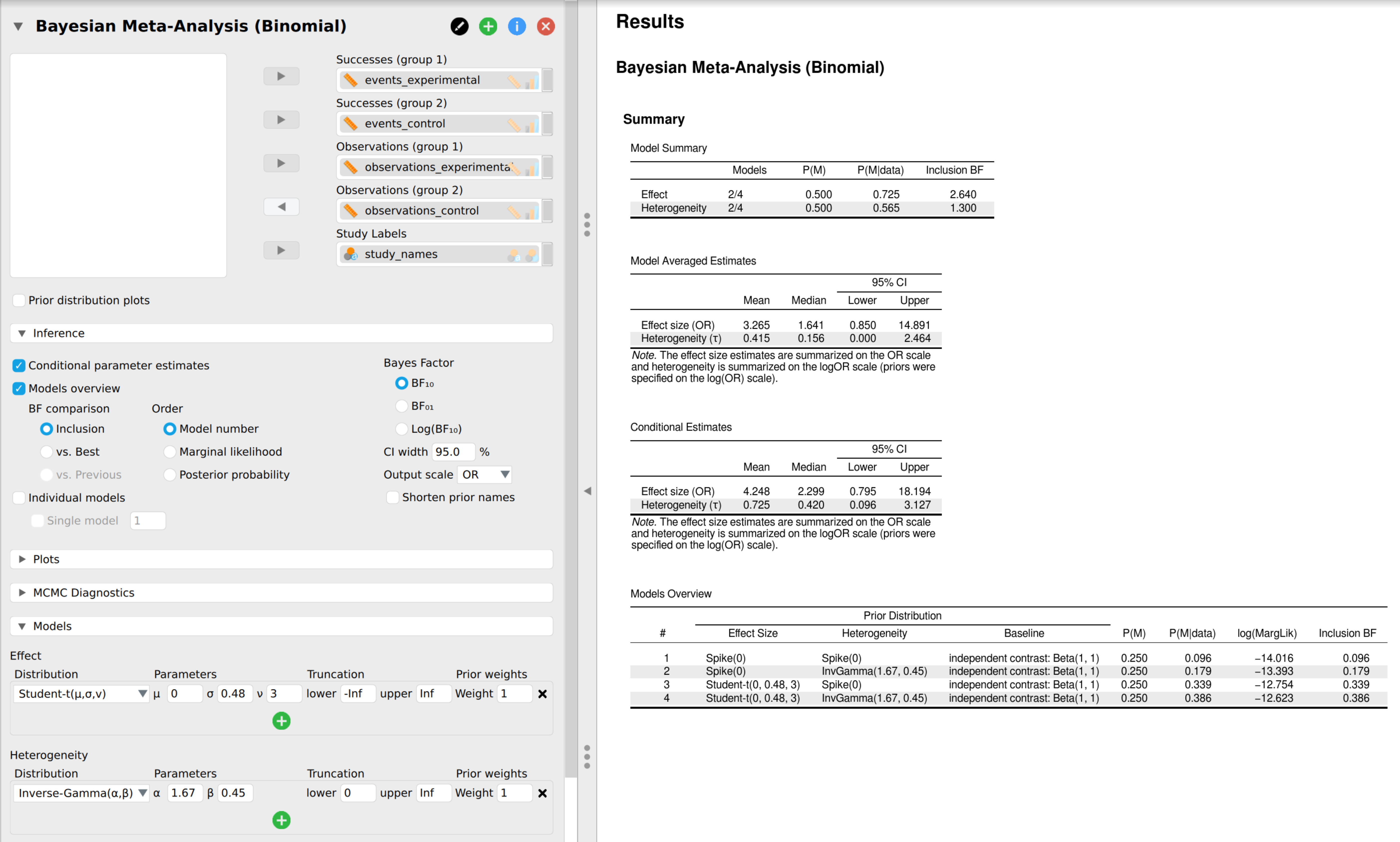}
    \caption{JASP screenshot of a Bayesian model-averaged meta-analysis of the Oduwole et al.\cite{oduwole2018honey} comparison concerning the everse events (nervousness, insomnia, hyperactivity) of honey in comparison to no treatment in treating acute cought in children. The left input panel shows the specification of the ``Acute Respiratory Infections'' CDSR subfield-specific prior distributions for logOR and heterogeneity $\tau$. The right output panel shows the corresponding results.}
    \label{fig:JASP_example}
\end{figure*}

Figure~\ref{fig:JASP_example} shows the JASP graphical user interface. The left panel shows settings for specifying the analysis settings and the right panel displaying the analysis output. The JASP output panel displays the corresponding BMA meta-analysis results. The ``Model Summary'' table summarizes the model-averaged evidence for the presence vs. absence of the effect and heterogeneity. The inclusion Bayes factors indicate weak evidence for the presence of the effect $\text{BF}_{10} = 2.64$ and virtualy no evidence for either the presence or absence of heterogeneity $\text{BF}_{\text{rf}} = 1.30$. The ``Conditional Estimates'' table summarizes conditional the meta-analytic estimates (the effect size estimate assuming presence of the effect and the heterogeneiry estimate assuming presence of heterogeneity). We find the effect size estimate $\text{OR} = 4.25$, 95\% [$0.80$, $18.20$] and heterogeneity estimate $\tau_{\text{logOR}} = 0.73$, 95\% [$0.10$, $3.13$]. We can notice two differences from the original frequentist output; 1) the effect size estimate is smaller and the credible interval is much narrower then previously reported, and 2) we are able to obtain a (very wide) estimate of the between-study heterogeneity. Both are a result of incorporating the existing information about the usual degrees of effect sizes in a given field which is especially relevant in cases with a very few observations.

The JASP meta-analytic module also offers a multitude of visualizationoptions (such a as forest plot, prior and posterior distributions, etc.) and additional options and analyses such as performing one-sided hypothesis tests, and various publication bias adjustment methods.\cite{bartos2020adjusting, berkhout2021tutorial}

\section{Concluding Comments}

In this article, we developed informed empirical prior distributions for the meta-analytic mean and heterogeneity parameter for meta-analyses of binary (log OR, log RR, and RD) and time-to-event (log HR) outcomes. We provided both general and topic-specific prior distributions based on almost 12,000 binary outcomes meta-analyses and general prior distributions based on 200 time-to-event meta-analyses extracted from the Cochrane Database of Systematic Reviews. Our results extend our previous work\cite{bartos2021bayesian} where we developed empirical prior distributions for continuous outcomes and the work of Pullenayegum\cite{pullenayegum2011informed} and Turner and colleagues\cite{turner2015predictive} who developed prior distributions for the heterogeneity parameter of log OR based on an earlier version of the database.

The newly developed prior distributions can be combined with Bayesian model-averaged meta-analysis to quantify the evidence in favor of or against the presence of the mean meta-analytic effect and heterogeneity. Moreover, Bayesian model-averaged meta-analysis does not force the analyst to base the entire inference on a single model. Instead, Bayesian model-averaged meta-analysis acknowledges the model uncertainty and combines the evidence and parameter estimates from the competing models according to their predictive performance.\cite{gronau2021primer, bartos2021bayesian} Both the direct quantification of evidence and accounting for model uncertainty is especially important in small sample sizes which are typical for medical meta-analyses, as it allows the analyst to disentangle the absence of evidence from the evidence of absence and avoid overconfident conclusions.\cite{keysers2020using, robinson2019what} 

Our results illustrate substantial uncertainty about the most appropriate meta-analytic model. No combination of the usual model types, the null vs the alternative hypothesis model and the fixed vs the random effects model, clearly dominated the other models across the examined meta-analyses. Moreover, the meta-analyses of binary outcomes were more in line with the null hypothesis and fixed effect models, while the opposite held true for meta-analyses of time-to-event outcomes. Although these results go against the common belief that the random-effects alternative model is the best-suited model for analysing data,\cite{higgins2009re,chung2013avoiding,council1992combining, mosteller1996understanding} the large uncertainly and considerable support for all model types echo our previous findings from meta-analyses of continuous outcomes.\cite{bartos2021bayesian}

We implemented the Bayesian binomial-normal meta-analytic model for log OR within the Bayesian model-averaged meta-analytic framework of the \texttt{RoBMA} \texttt{R} package\cite{RoBMA3} (alongside the already existing normal-normal meta-analytic model). We incorporated the framework and empirical prior distributions to into the user-friendly graphical user interface of JASP.\cite{JASP17, ly2021bayesian} Finally, we illustrated the programs on and applied the empirical prior distributions to an example of acute respiratory infections.

\section*{Acknowledgments}
This work was supported by The Netherlands Organisation for Scientific Research (NWO) through a Research Talent grant (to QFG; 406.16.528), a Vici grant (to EJW; 016.Vici.170.083), and an NWA Idea Generator grant (to WMO; NWA.1228.191.045).

\section*{Financial disclosure}
None reported.

\section*{Data availability statement}
Data and analysis scripts are publicly available at: \url{https://osf.io/v9bj6/}.

\section*{Conflict of interest}
František Bartoš, Alexander Ly, and Eric-Jan Wagenmakers declare their involvement in the open-source software package JASP (\url{https://jasp-stats.org}), a non-commercial, publicly-funded effort to make Bayesian statistics accessible to a broader group of researchers and students.
Willem Otte is co-founder of RCTAlert (\url{https://rctalert.com}), a commercial AI platform providing weekly clinical trial notifications.

\bibliographystyle{WileyNJD-AMA}
\bibliography{manuscript.bib}

\newpage
\appendix

\section{R Code for the Adverse effects of honey in treating acute cough in children Example}
\label{app:example_R}

This Appendix shows how to conduct the example analysis with the statistical programming language \texttt{R}.\cite{R} First, we need to install the \texttt{RoBMA} \texttt{R} package,\cite{RoBMA3} (this command needs to be executed only if the package is not already installed):
\begin{verbatim}
install.packages("RoBMA")
\end{verbatim}

\noindent After the \texttt{RoBMA} package has been installed, we load it into the \texttt{R} session,
\begin{verbatim}
library("RoBMA")
\end{verbatim}
\noindent and specify the number of events and observations in both the experimental and control group from both studies,
\begin{verbatim}
events_experimental        <- c(5, 2)
events_control             <- c(0, 0)
observations_experimental  <- c(35, 40)
observations_control       <- c(39, 40)
study_names <- c("Paul 2007", "Shadkam 2010")
\end{verbatim}

\noindent In order to use the binomial-normal BMA meta-analysis we use the \texttt{BiBMA()} function. We specify the number of of events and observations in each group as the corresponding input (\texttt{x1}, \texttt{x2}, \texttt{n1}, \texttt{n2}) and set the subfield-specific prior distributions using the \texttt{priors\_effect} and \texttt{priors\_heterogeneity} arguments accordingly to the ``Acute Respiratory Infections'' row in Table~\ref{tab:priors-log-OR}:  
 
\begin{verbatim}
fit <- BiBMA(
  x1 = events_experimental,
  x2 = events_control,
  n1 = observations_experimental,
  n2 = observations_control,
  study_names = study_names,
  priors_effect        = prior(distribution="t", parameters=list(location=0, scale=0.48, df=3)),
  priors_heterogeneity = prior(distribution="invgamma",  parameters=list(shape=1.67, scale=0.45)),
  seed = 1)
\end{verbatim}

\noindent To obtain the inclusion Bayes factors and conditional posterior distribution summaries, we use the \texttt{summary} function with the \texttt{conditional = TRUE} argument,
\begin{verbatim}
summary(fit, conditional = TRUE, output_scale = "OR")
\end{verbatim}

\noindent which produces output corresponds to that given by JASP (up to MCMC error):
\begin{verbatim}
> summary(fit, conditional = TRUE, output_scale = "OR")
Call:
BiBMA(x1 = events_experimental, x2 = events_control, n1 = observations_experimental, 
    n2 = observations_control, study_names = study_names, priors_effect = prior(distribution = "t", 
        parameters = list(location = 0, scale = 0.48, df = 3)), 
    priors_heterogeneity = prior(distribution = "invgamma", parameters = list(shape = 1.67, 
        scale = 0.45)), seed = 1)

Bayesian model-averaged meta-analysis (binomial-normal model)
Components summary:
              Models Prior prob. Post. prob. Inclusion BF
Effect           2/4       0.500       0.725        2.630
Heterogeneity    2/4       0.500       0.564        1.296

Model-averaged estimates:
     Mean Median 0.025  0.975
mu  3.389  1.642 0.842 15.143
tau 0.420  0.158 0.000  2.594
The effect size estimates are summarized on the OR scale and heterogeneity is summarized 
on the logOR scale (priors were specified on the log(OR) scale).

Conditional estimates:
     Mean Median 0.025  0.975
mu  4.242  2.261 0.781 17.613
tau 0.747  0.426 0.097  3.233
The effect size estimates are summarized on the OR scale and heterogeneity is summarized 
on the logOR scale (priors were specified on the log(OR) scale).
\end{verbatim}

\section{Analysis of log RR and RD}
\label{app:other}

This Appendix contains tables and figures for the log RR and RD.

\begin{figure}[ht!]
\begin{tabular}{cc}
    \begin{minipage}{.50 \textwidth}
    \includegraphics[width=1\textwidth]{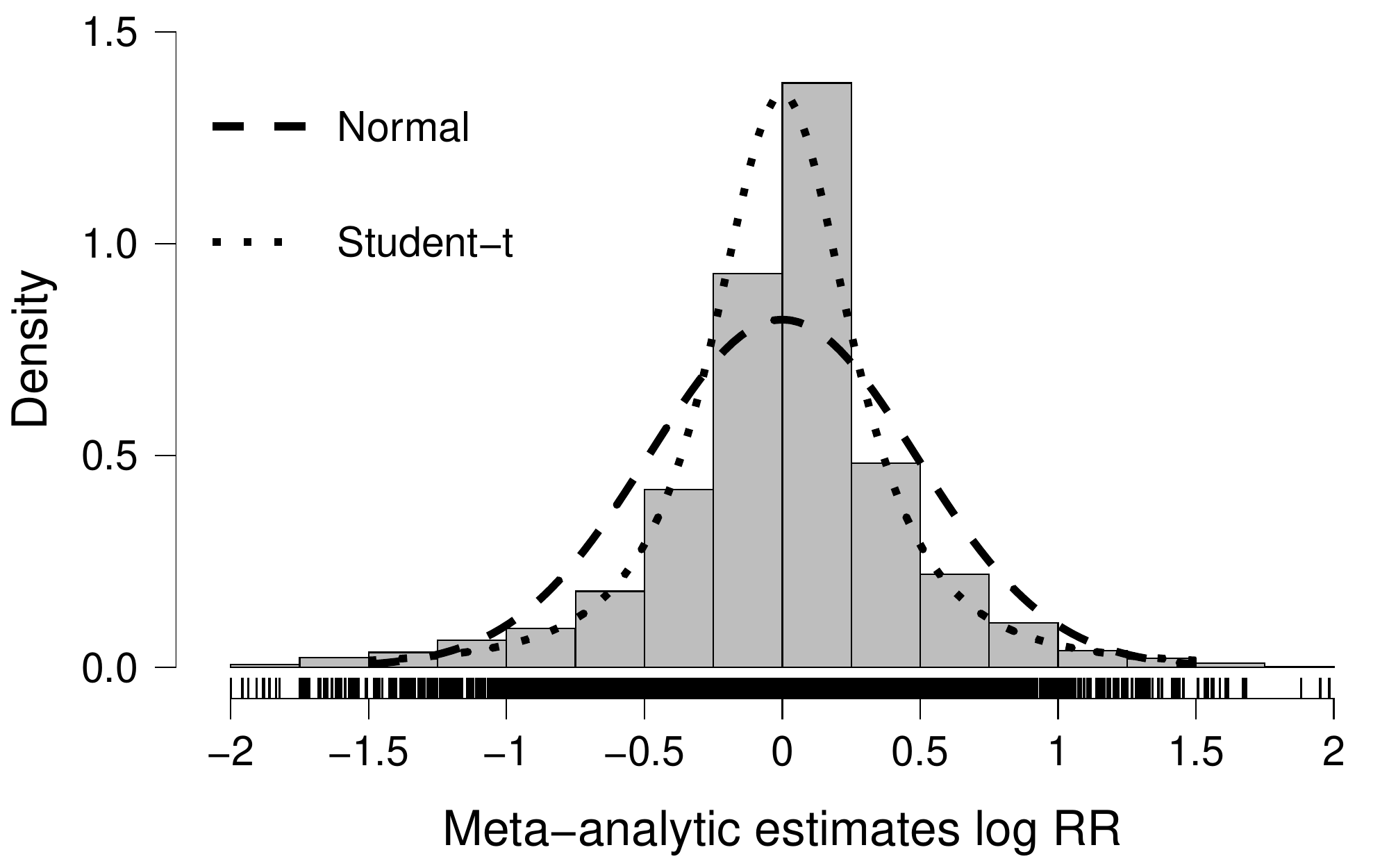}
    \end{minipage}  &
    \begin{minipage}{.50 \textwidth}
    \includegraphics[width=1\textwidth]{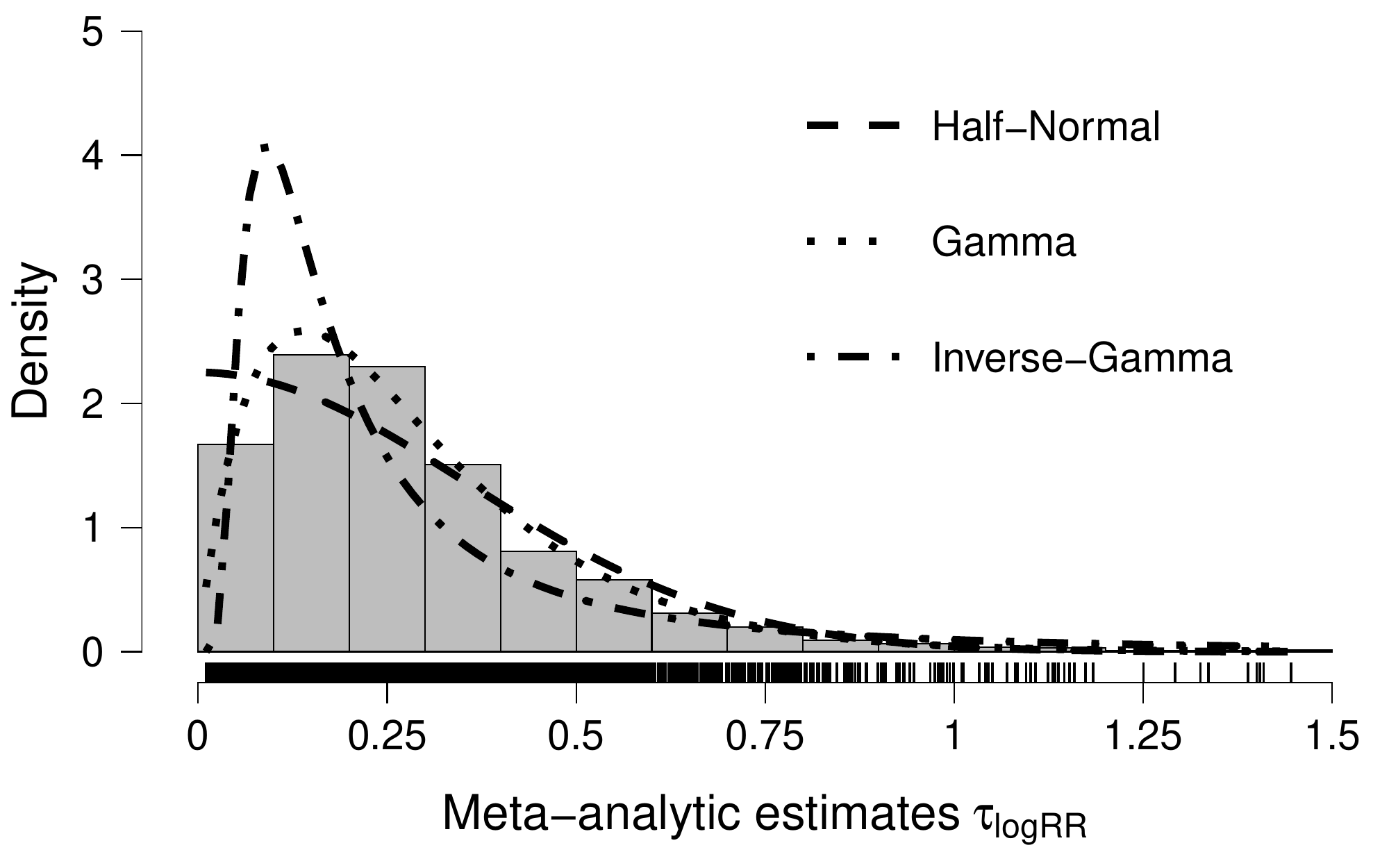}
    \end{minipage} 
    \\
    \begin{minipage}{.50 \textwidth}
    \includegraphics[width=1\textwidth]{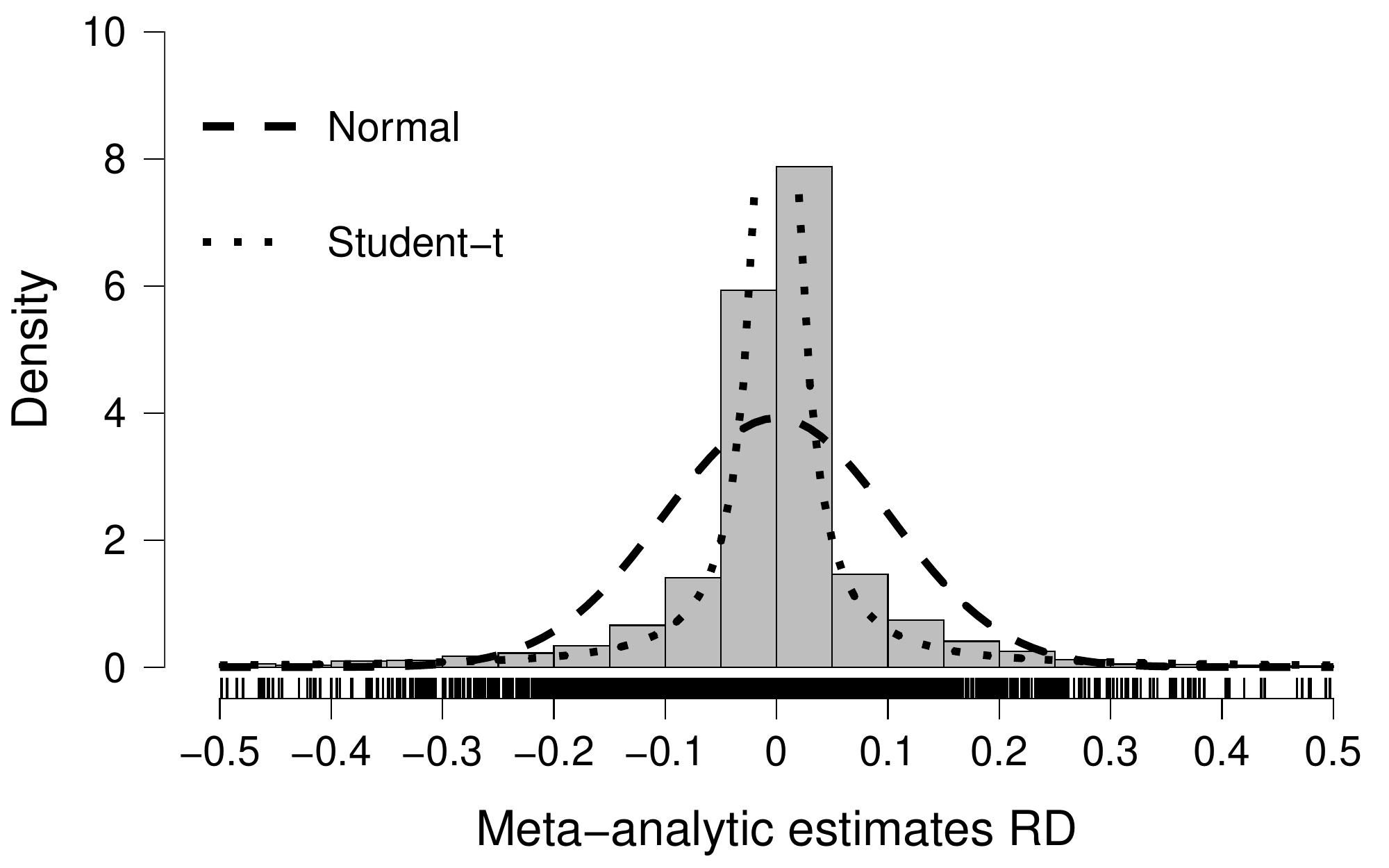}
    \end{minipage}  & 
    \begin{minipage}{.50 \textwidth}
    \includegraphics[width=1\textwidth]{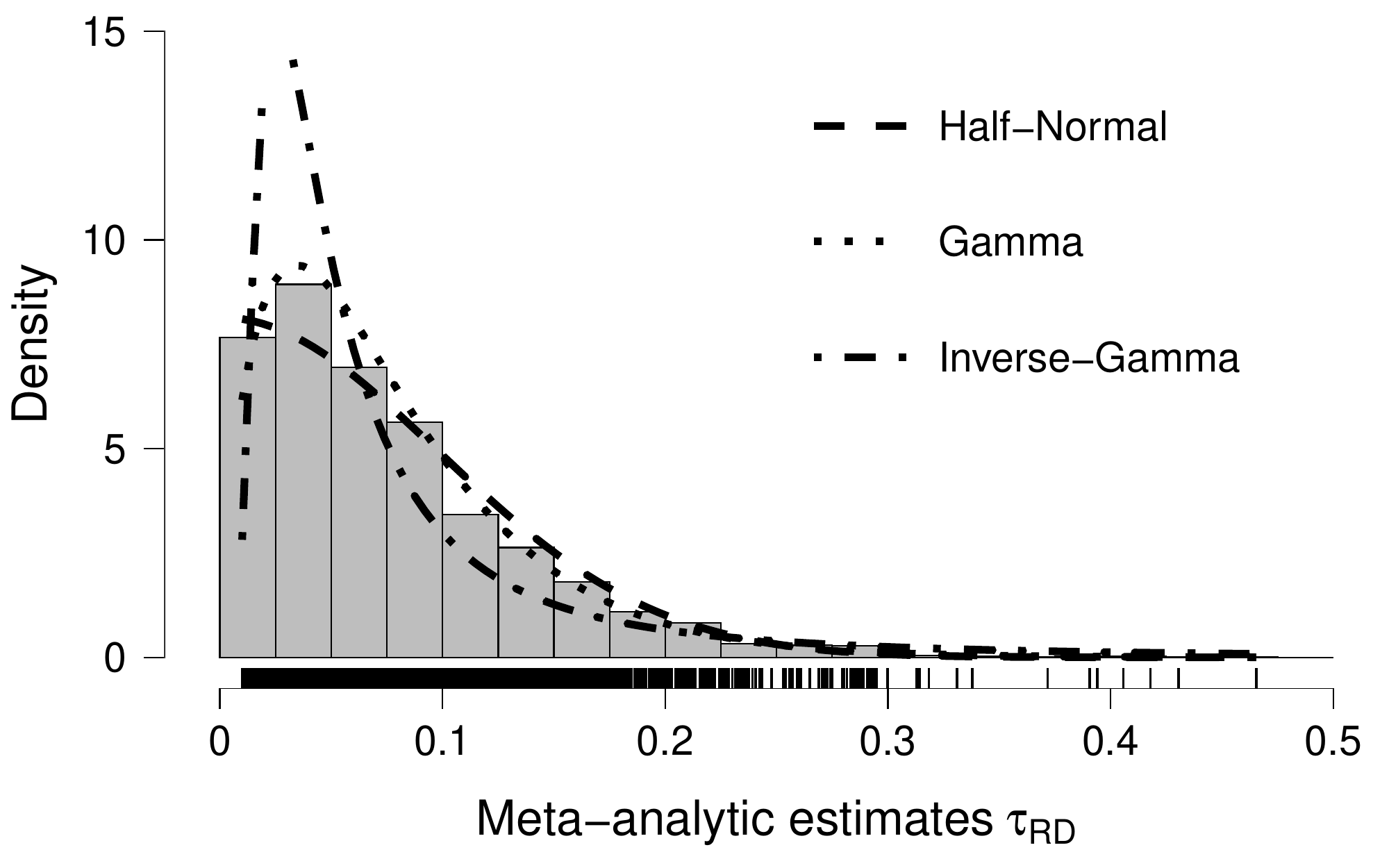}
    \end{minipage} 
    \\
\end{tabular}
\caption{Frequentist effect sizes estimates and candidate prior distributions from the training data set. Binary outcomes as log RR (first row) and RD (second row). Histogram and tick marks display the estimated effect size estimates (left) and between-study standard deviation estimates (right), whereas lines represent candidate prior distributions for the population effect size parameter (left) and candidate prior distributions for the population between-study standard deviation $\tau$ (right; see Table~\ref{tab:prior-training-dataset}). 34 log RR outside of the $\pm 2$ range are not shown and 31 RD outside the $\pm 0.5$ range are not shown. 6 $\tau$ estimates larger than 1.5 and 2,401 $\tau$ estimates lower than 0.01 of log RR are not shown. 2,823 $\tau$ estimates lower than 0.01 of RD are not shown.}
\label{fig:prior-training-dataset2}
\end{figure}

\begin{table*}
  \centering
  \caption{Candidate prior distributions for the $\delta$ and $\tau$ parameters as obtained from the training set. The Student's $t$-distributions parameterized by location, scale, and degrees of freedom, the (half) normal distributions parameterized by mean and standard deviation, and the inverse-gamma and gamma distributions parameterized by shape and scale. See Figure~\ref{fig:prior-training-dataset2}.}
  \label{tab:prior-training-dataset2}
    \begin{tabular}{ll}
    \multicolumn{2}{l}{Binary outcomes as log RR} \\
    \midrule
    $\mu \sim \mathcal{N}(0, 0.49)$           & $\tau \sim \mathcal{N}_+(0, 0.35)$         \\
    $\mu \sim \mathcal{T}(0, 0.26, 2.28)$     & $\tau \sim \text{Inv-Gamma}(1.51, 0.23)$   \\
                                              & $\tau \sim \text{Gamma}(1.96, 0.14)$       \\ 
    \bottomrule \\
    \multicolumn{2}{l}{Binary outcomes as RD} \\
    \midrule
    $\mu \sim \mathcal{N}(0, 0.10)$         & $\tau \sim \mathcal{N}_+(0, 0.10)$        \\
    $\mu \sim \mathcal{T}(0, 0.02, 0.85)$   & $\tau \sim \text{Inv-Gamma}(1.68, 0.07)$  \\
                                            & $\tau \sim \text{Gamma}(1.80, 0.04)$      \\ 
    \bottomrule \\
  \end{tabular}
\end{table*}

\begin{table*}
    \centering
    \caption{Ranking totals for each prior distribution in $\mathcal{H}_1^r$ based on the test set. The numbers indicate how many times a specific prior distribution attained a specific posterior probability rank. Rank `1' represents the best performance. The rankings reflect predictive adequacy that is model-averaged across the possible prior distribution configurations of the other parameter.}
    \label{tab:ranking-priors2}
    \begin{tabular}{@{}lrrrrr@{}}
        \multicolumn{6}{l}{Binary outcomes as log RR} \\
        \midrule
                                                  & \multicolumn{3}{c}{Rank}  &       &             \\
        Prior distribution                        &    1 &    2 &    3        & PrMP* & AV. PoMP**  \\
        \\
        \multicolumn{6}{l}{Parameter $\mu$}                                                      \\
        \midrule
        $\text{Normal}(0, 0.49)$          &  6817 & 14965 & --- & 0.50 & 0.48 \\ 
        $\text{Student-t}(0, 0.26, 2.28)$ & 14965 &  6817 & --- & 0.50 & 0.52 \\
        \\
        \multicolumn{6}{l}{Parameter $\tau$} \\
        \midrule      
        $\text{Normal}_+(0, 0.35)$     & 4311 & 12760 &  4711 & 0.33 & 0.34 \\ 
        $\text{Inv-Gamma}(1.51, 0.23)$ & 8151 &  1965 & 11666 & 0.33 & 0.33 \\ 
        $\text{Gamma}(1.96, 0.14)$     & 9320 &  7057 &  5405 & 0.33 & 0.33 \\ 
        \bottomrule
        \\ 
        \multicolumn{6}{l}{Time to event outcomes as RD} \\
        \midrule
                                                  & \multicolumn{3}{c}{Rank}  &       &             \\
        Prior distribution                        &    1 &    2 &    3        & PrMP* & AV. PoMP**  \\
        \\
        \multicolumn{6}{l}{Parameter $\mu$} \\
        \midrule
        $\text{Normal}(0, 0.10)$          &  6403 & 15379 & --- & 0.50 & 0.42 \\ 
        $\text{Student-t}(0, 0.02, 0.85)$ & 15379 &  6403 & --- & 0.50 & 0.58 \\ 
        \\
        \multicolumn{6}{l}{Parameter $\tau$} \\
        \midrule
        $\text{Normal}_+(0, 0.10)$    &  8355 &  6987 & 6440 & 0.33 & 0.37 \\ 
        $\text{InvGamma}(1.68, 0.07)$ & 10340 &  2954 & 8488 & 0.33 & 0.31 \\ 
        $\text{Gamma}(1.8, 23.32)$    &  3087 & 11841 & 6854 & 0.33 & 0.32 \\ 
        \bottomrule
        
        \multicolumn{6}{l}{*Prior model probability} \\
        \multicolumn{6}{l}{**Average posterior model probability} \\
    \end{tabular}
\end{table*}

\begin{figure}
\begin{tabular}{cc}
    \begin{minipage}{.50 \textwidth}
    \includegraphics[width=1\textwidth]{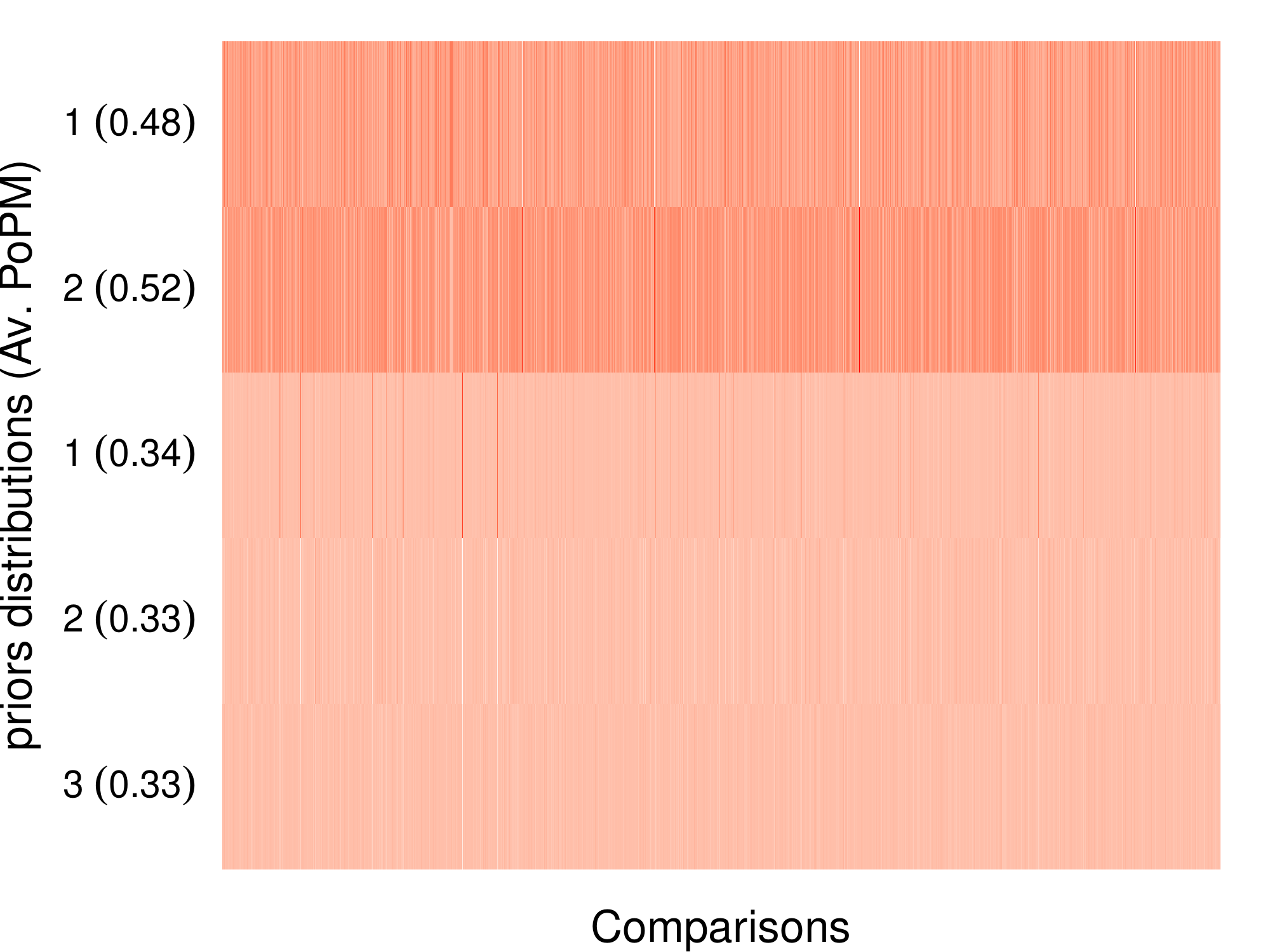}
    \end{minipage}  &
    \begin{minipage}{.50 \textwidth}
    \includegraphics[width=1\textwidth]{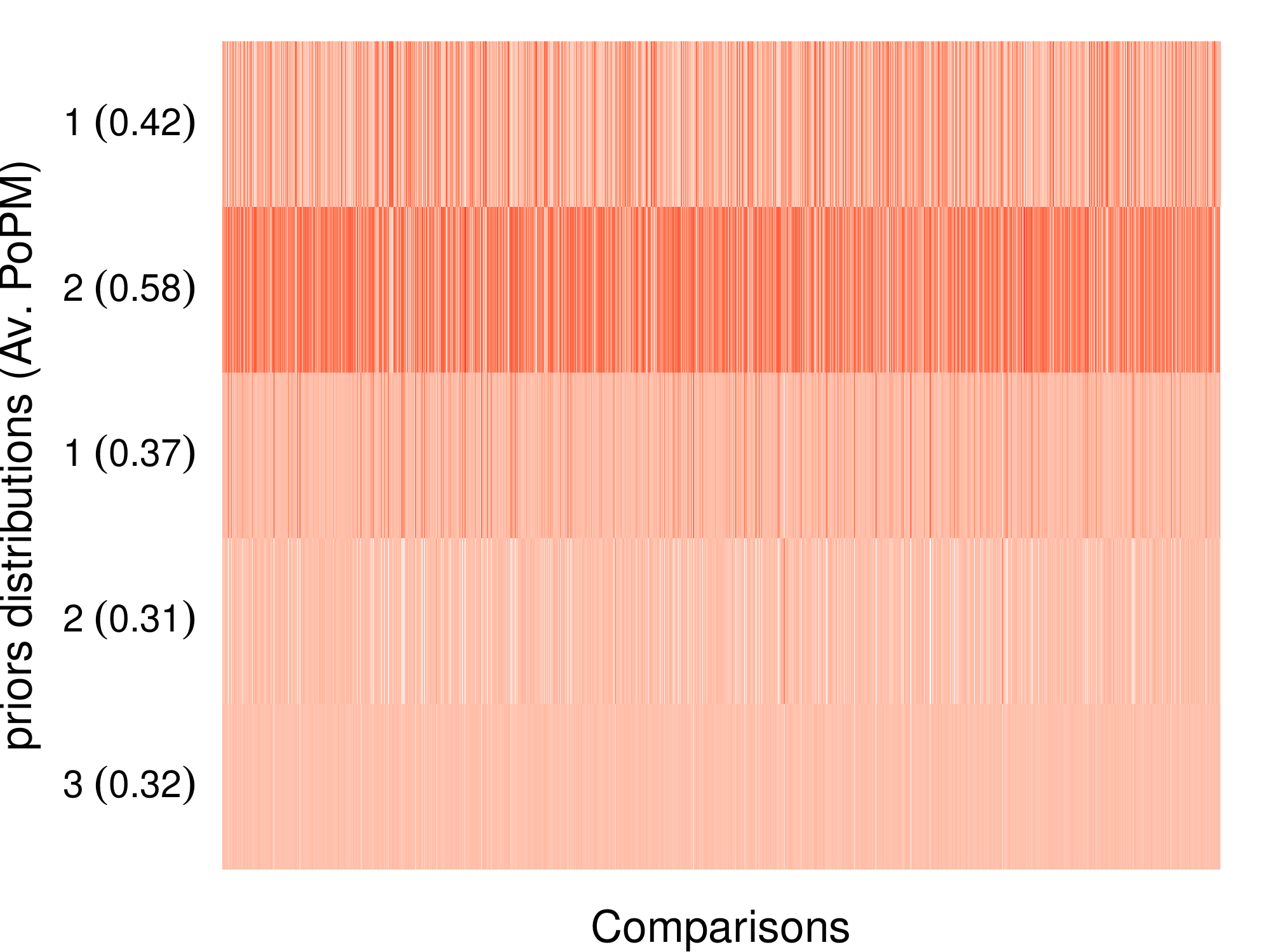}
    \end{minipage} 
\end{tabular}
\caption{Posterior model probability of the competing prior distributions for log RR and RD. For each comparison, the color gradient ranges from white (low posterior probability) to dark red (high posterior probability). The numbers in parentheses are the averaged posterior probabilities across all comparisons.}
\label{fig:posterior-priors2}
\end{figure}

\begin{table*}
    \centering
    \caption{Ranking totals for each model type in the test set. The numbers indicate how many times a specific model type attained a specific posterior probability rank. Rank `1' represents the best performance. The rankings reflect predictive adequacy that is model-averaged across the possible prior distribution configurations.}
    \label{tab:ranking-models2}
    \begin{tabular}{@{}lrrrrrr@{}}

        \multicolumn{6}{l}{Binary outcomes as log RR} \\
        \midrule
                             & \multicolumn{4}{c}{Rank}       &       &             \\
        Model                &      1 &     2 &     3 &     4 & PrMP* & AV. PoMP**  \\ 
        \midrule
        $\mathcal{H}_0^f$    &   9604 & 1650 & 1776 & 8748 & 0.25 & 0.24 \\ 
        $\mathcal{H}_1^f$    &   5557 & 4305 & 9815 & 2101 & 0.25 & 0.26 \\ 
        $\mathcal{H}_0^r$    &   2348 & 9882 & 8292 & 1260 & 0.25 & 0.21 \\ 
        $\mathcal{H}_1^r$    &   4273 & 5945 & 1895 & 9669 & 0.25 & 0.29 \\ 
        \bottomrule
        \\
        \multicolumn{6}{l}{Binary outcomes as RD} \\
        \midrule
                             & \multicolumn{4}{c}{Rank}       &       &     \\
        Model                &    1 &   2 &   3 &   4 & PrMP* & AV. PoMP**  \\
        \midrule
        $\mathcal{H}_0^f$    &   10644 &  1449 &  1632 &  8027 & 0.25 & 0.33 \\ 
        $\mathcal{H}_1^f$    &    3683 & 10165 &  5924 &  1980 & 0.25 & 0.22 \\ 
        $\mathcal{H}_0^r$    &    2385 &  5892 & 12774 &   731 & 0.25 & 0.18 \\ 
        $\mathcal{H}_1^r$    &    5070 &  4276 &  1422 & 11014 & 0.25 & 0.27 \\ 
        \bottomrule
  
        \multicolumn{7}{l}{*Prior model probability} \\
        \multicolumn{7}{l}{**Average posterior model probability} \\
    \end{tabular}
\end{table*}

\begin{figure}
\begin{tabular}{cc}
    \begin{minipage}{.50 \textwidth}
    \includegraphics[width=1\textwidth]{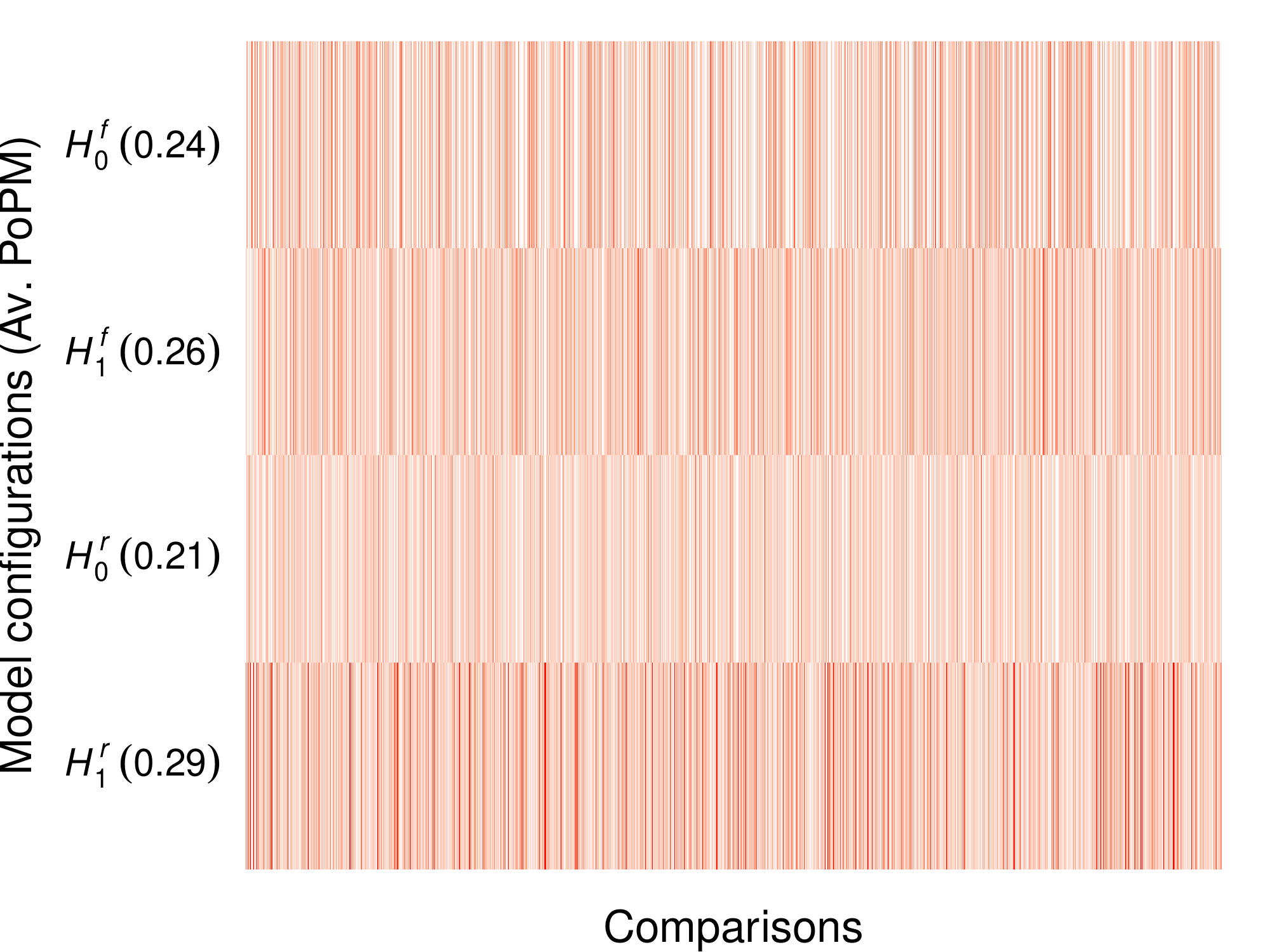}
    \end{minipage}  &
    \begin{minipage}{.50 \textwidth}
    \includegraphics[width=1\textwidth]{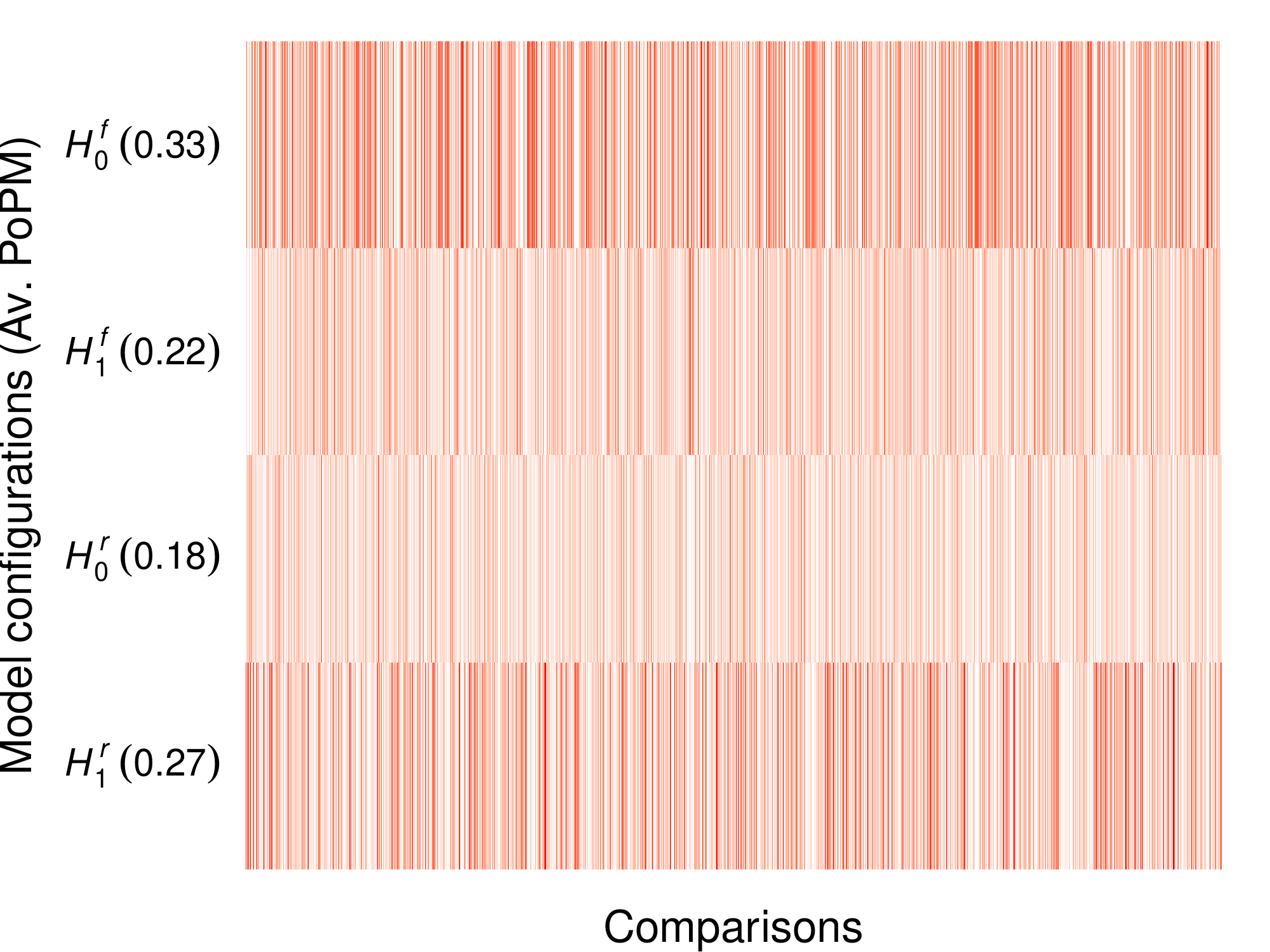}
    \end{minipage} 
\end{tabular}
\caption{Posterior model probability of the competing prior model types for log OR and log HR. For each comparison, the color gradient ranges from white (low posterior probability) to dark red (high posterior probability). The numbers in parentheses are the averaged posterior probabilities across all comparisons.}
\label{fig:posterior-hypotheses2}
\end{figure}

\begin{figure}
\begin{tabular}{cc}
    \begin{minipage}{.50 \textwidth}
    \includegraphics[width=1\textwidth]{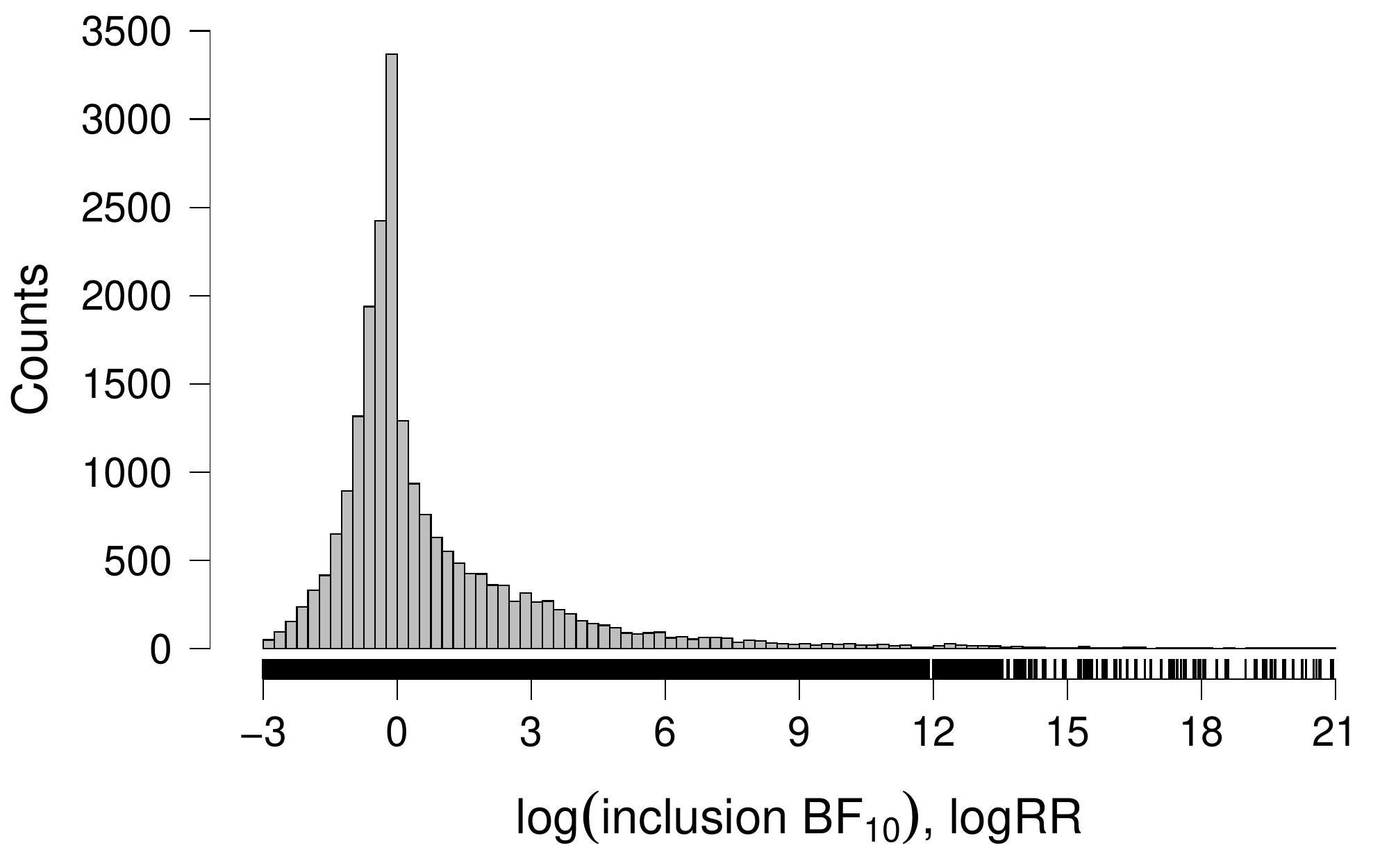}
    \end{minipage}  &
    \begin{minipage}{.50 \textwidth}
    \includegraphics[width=1\textwidth]{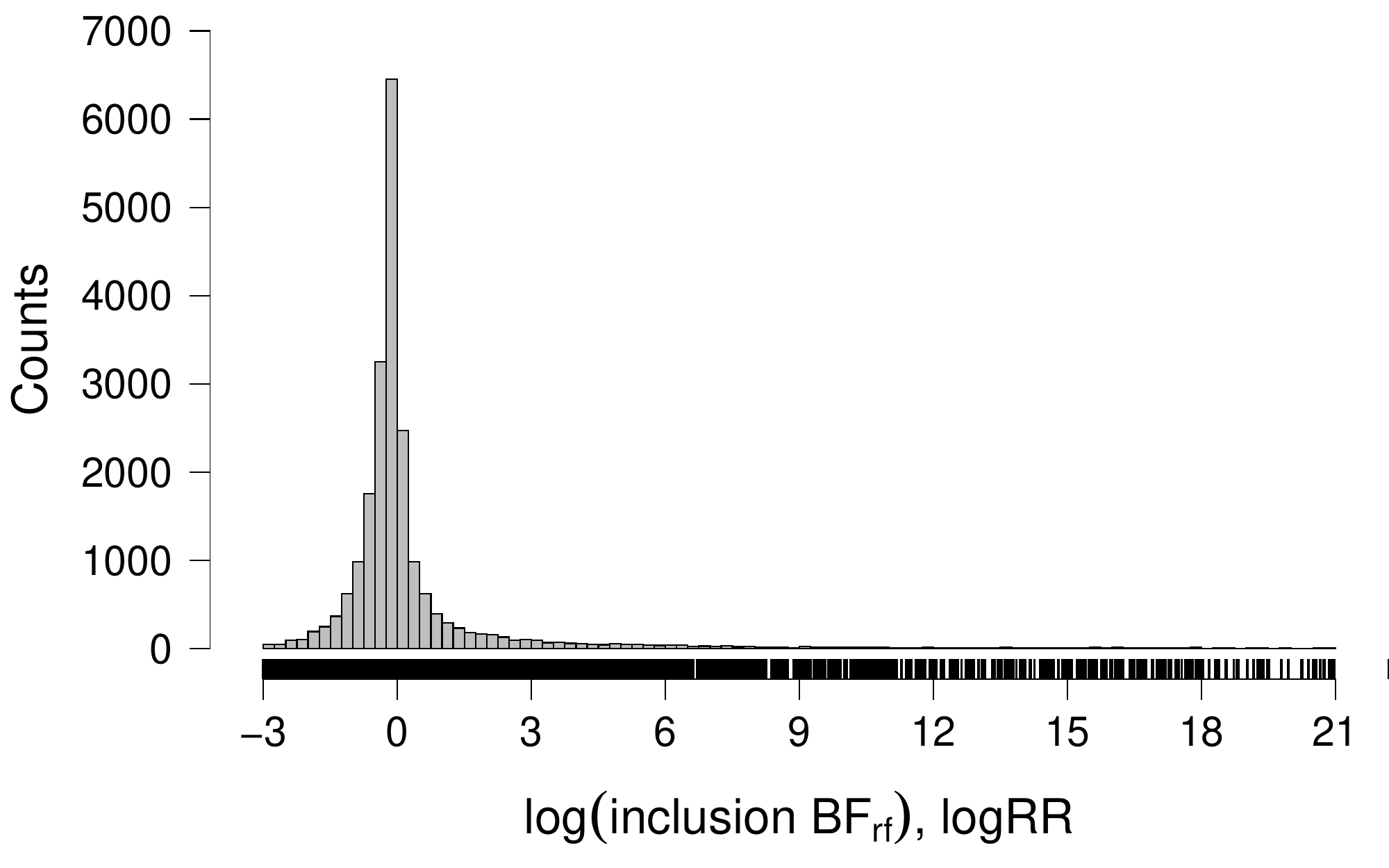}
    \end{minipage}  
    \\
    \begin{minipage}{.50 \textwidth}
    \includegraphics[width=1\textwidth]{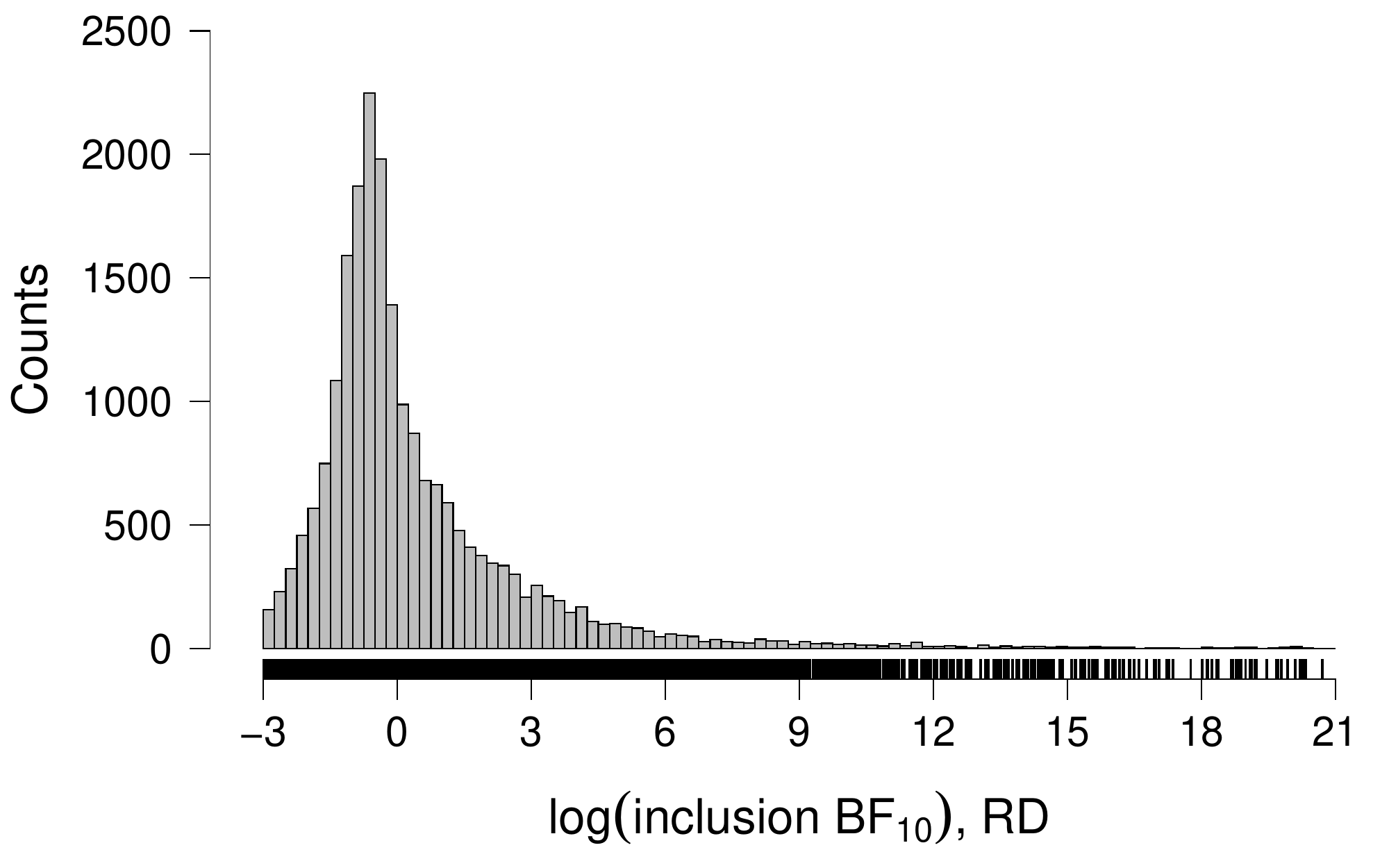}
    \end{minipage}  &
    \begin{minipage}{.50 \textwidth}
    \includegraphics[width=1\textwidth]{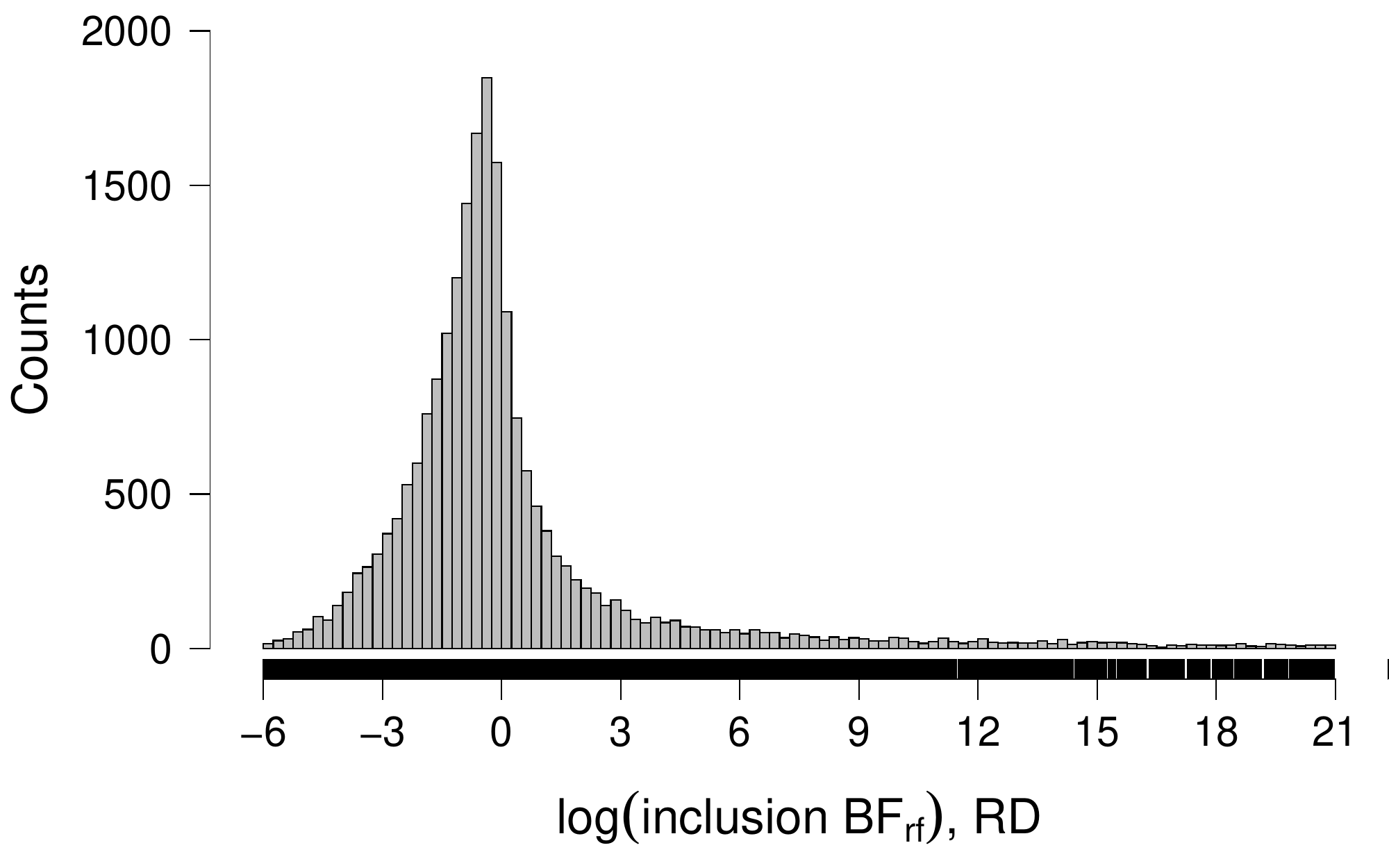}
    \end{minipage}  
\end{tabular}
\caption{Inclusion Bayes factors in favor of the presence of a treatment effect (left) and in favor of the presence of across-study heterogeneity (right) in the test set. log RR (first row) and RD (second row). In log RR; for the presence of the effect, 140 log Bayes factors larger than 21 and 91 log Bayes factors lower than -3, and for the presence of heterogeneity, 387 log Bayes factors larger than 21 and 83 log Bayes factors lower than -3 are not shown. In RD; for the presence of the effect, 123 log Bayes factors larger than 21 and 386 log Bayes factors lower than -3, and for the presence of heterogeneity, 902 log Bayes factors larger than 21 and 41 log Bayes factors lower than -6 are not shown.}
\label{fig:inclusion-BF2}
\end{figure}

\begin{table}
\centering
\caption{General and subfield-specific prior distributions for log RR of individual topics from the Cochrane database of systematic reviews estimated by hierarchical regression based on the complete data set. The Student's $t$-distribution parameterized by location, scale, and degrees of freedom and the Inverse-Gamma distribution parameterized by shape and scale.}
\label{tab:priors-log-RR}
\small
\begin{tabular}{lrrrr}
Topic & Comparisons $\mu$ & Comparisons $\tau$ & Prior $\mu$ & Prior $\tau$ \\
\toprule
Acute Respiratory Infections & 299 (5721) & 178 (3806) & Student-t(0, 0.27, 3) & Inv-Gamma(1.58, 0.25) \\ 
  Airways & 449 (8806) & 207 (3723) & Student-t(0, 0.25, 3) & Inv-Gamma(1.36, 0.15) \\ 
  Anaesthesia & 409 (9471) & 231 (6158) & Student-t(0, 0.45, 5) & Inv-Gamma(1.21, 0.21) \\ 
  Back and Neck & 14 (295) & 8 (186) & Student-t(0, 0.34, 3) & Inv-Gamma(1.52, 0.22) \\ 
  Bone, Joint and Muscle Trauma & 198 (3616) & 86 (1455) & Student-t(0, 0.26, 2) & Inv-Gamma(0.97, 0.14) \\ 
  Breast Cancer & 136 (2622) & 115 (2291) & Student-t(0, 0.21, 3) & Inv-Gamma(1.49, 0.24) \\ 
  Childhood Cancer & 1 (13) & --- & Student-t(0, 0.28, 3) & --- \\ 
  Colorectal & 235 (4157) & 133 (2523) & Student-t(0, 0.38, 4) & Inv-Gamma(1.42, 0.27) \\ 
  Common Mental Disorders & 640 (13175) & 425 (9704) & Student-t(0, 0.29, 3) & Inv-Gamma(1.43, 0.20) \\ 
  Consumers and Communication & 44 (918) & 40 (865) & Student-t(0, 0.13, 2) & Inv-Gamma(1.48, 0.14) \\ 
  Cystic Fibrosis and Genetic Disorders & 76 (1692) & 31 (736) & Student-t(0, 0.21, 2) & Inv-Gamma(1.71, 0.24) \\ 
  Dementia and Cognitive Improvement & 124 (2047) & 72 (1199) & Student-t(0, 0.29, 3) & Inv-Gamma(1.53, 0.19) \\ 
  Developmental, Psychosocial and Learning Problems & 107 (1467) & 85 (1162) & Student-t(0, 0.39, 3) & Inv-Gamma(0.97, 0.11) \\ 
  Drugs and Alcohol & 105 (2019) & 66 (1376) & Student-t(0, 0.22, 4) & Inv-Gamma(1.58, 0.20) \\ 
  Effective Practice and Organisation of Care & 58 (1093) & 46 (904) & Student-t(0, 0.30, 4) & Inv-Gamma(1.40, 0.24) \\ 
  Emergency and Critical Care & 232 (4776) & 145 (3438) & Student-t(0, 0.19, 2) & Inv-Gamma(1.65, 0.25) \\ 
  ENT & 16 (233) & 5 (63) & Student-t(0, 0.40, 4) & Inv-Gamma(1.50, 0.24) \\ 
  Epilepsy & 116 (2064) & 49 (1090) & Student-t(0, 0.58, 5) & Inv-Gamma(1.74, 0.27) \\ 
  Eyes and Vision & 57 (984) & 38 (719) & Student-t(0, 0.34, 4) & Inv-Gamma(1.60, 0.26) \\ 
  Fertility Regulation & 60 (1032) & 42 (744) & Student-t(0, 0.34, 4) & Inv-Gamma(1.59, 0.31) \\ 
  Gut & 308 (5941) & 202 (4348) & Student-t(0, 0.40, 4) & Inv-Gamma(1.69, 0.29) \\ 
  Gynaecological, Neuro-oncology and Orphan Cancer & 242 (5784) & 154 (3988) & Student-t(0, 0.35, 4) & Inv-Gamma(1.43, 0.25) \\ 
  Gynaecology and Fertility & 373 (6967) & 181 (3666) & Student-t(0, 0.25, 2) & Inv-Gamma(1.29, 0.16) \\ 
  Haematology & 153 (4472) & 95 (2958) & Student-t(0, 0.31, 3) & Inv-Gamma(1.22, 0.11) \\ 
  Heart & 950 (19836) & 604 (13252) & Student-t(0, 0.17, 3) & Inv-Gamma(1.96, 0.24) \\ 
  Heart; Vascular & 8 (178) & 7 (166) & Student-t(0, 0.39, 3) & Inv-Gamma(1.62, 0.25) \\ 
  Hepato-Biliary & 1037 (36554) & 578 (21332) & Student-t(0, 0.23, 3) & Inv-Gamma(1.44, 0.19) \\ 
  HIV/AIDS & 33 (486) & 17 (228) & Student-t(0, 0.21, 3) & Inv-Gamma(1.66, 0.23) \\ 
  Hypertension & 66 (1127) & 33 (515) & Student-t(0, 0.25, 4) & Inv-Gamma(1.48, 0.11) \\ 
  Incontinence & 78 (1479) & 54 (1020) & Student-t(0, 0.39, 3) & Inv-Gamma(1.53, 0.33) \\ 
  Infectious Diseases & 358 (6490) & 254 (4813) & Student-t(0, 0.41, 3) & Inv-Gamma(1.30, 0.27) \\ 
  Injuries & 214 (5639) & 138 (3839) & Student-t(0, 0.27, 4) & Inv-Gamma(1.46, 0.19) \\ 
  Kidney and Transplant & 474 (8737) & 275 (5205) & Student-t(0, 0.35, 4) & Inv-Gamma(1.51, 0.23) \\ 
  Lung Cancer & 75 (1402) & 64 (1131) & Student-t(0, 0.38, 4) & Inv-Gamma(1.49, 0.26) \\ 
  Metabolic and Endocrine Disorders & 131 (3233) & 72 (1768) & Student-t(0, 0.13, 1) & Inv-Gamma(1.55, 0.19) \\ 
  Methodology & 74 (2098) & 73 (2084) & Student-t(0, 0.27, 5) & Inv-Gamma(1.74, 0.23) \\ 
  Movement Disorders & 58 (1042) & 40 (765) & Student-t(0, 0.46, 4) & Inv-Gamma(1.54, 0.24) \\ 
  Multiple Sclerosis and Rare Diseases of the CNS & 28 (544) & 22 (467) & Student-t(0, 0.44, 3) & Inv-Gamma(1.58, 0.35) \\ 
  Musculoskeletal & 138 (2393) & 84 (1541) & Student-t(0, 0.33, 3) & Inv-Gamma(1.54, 0.31) \\ 
  Neonatal & 335 (6207) & 151 (2517) & Student-t(0, 0.18, 3) & Inv-Gamma(1.89, 0.30) \\ 
  Neuromuscular & 40 (647) & 18 (273) & Student-t(0, 0.41, 4) & Inv-Gamma(1.44, 0.15) \\ 
  Oral Health & 66 (1032) & 32 (465) & Student-t(0, 0.55, 4) & Inv-Gamma(1.50, 0.37) \\ 
  Pain, Palliative and Supportive Care & 310 (5965) & 212 (4298) & Student-t(0, 0.59, 5) & Inv-Gamma(1.98, 0.39) \\ 
  Pregnancy and Childbirth & 1300 (24223) & 813 (16138) & Student-t(0, 0.24, 2) & Inv-Gamma(1.40, 0.21) \\ 
  Schizophrenia & 601 (13147) & 347 (8290) & Student-t(0, 0.33, 3) & Inv-Gamma(1.47, 0.26) \\ 
  Sexually Transmitted Infections & 25 (318) & 18 (225) & Student-t(0, 0.39, 3) & Inv-Gamma(1.43, 0.21) \\ 
  Skin & 142 (2285) & 85 (1434) & Student-t(0, 0.41, 1) & Inv-Gamma(1.49, 0.23) \\ 
  Stroke & 301 (5363) & 149 (2476) & Student-t(0, 0.13, 2) & Inv-Gamma(1.52, 0.15) \\ 
  Tobacco Addiction & 215 (5018) & 174 (4276) & Student-t(0, 0.36, 5) & Inv-Gamma(2.02, 0.36) \\ 
  Urology & 89 (1694) & 62 (1309) & Student-t(0, 0.44, 3) & Inv-Gamma(1.37, 0.18) \\ 
  Vascular & 176 (2473) & 96 (1383) & Student-t(0, 0.43, 5) & Inv-Gamma(1.26, 0.19) \\ 
  Work & 14 (208) & 10 (166) & Student-t(0, 0.23, 3) & Inv-Gamma(1.50, 0.26) \\ 
  Wounds & 74 (1148) & 43 (688) & Student-t(0, 0.37, 4) & Inv-Gamma(1.63, 0.38) \\ 
  \bottomrule
  Pooled Estimate & 11862 (250331) & 7159 (159166) & Student-t(0, 0.32, 3) & Inv-Gamma(1.51, 0.23) \\ 
\bottomrule
\end{tabular}
\end{table}

\begin{table}
\centering
\caption{General and subfield-specific prior distributions for RD of individual topics from the Cochrane database of systematic reviews estimated by hierarchical regression based on the complete data set. The Student's $t$-distribution parameterized by location, scale, and degrees of freedom and the Half-Normal distribution parameterized by mean and standard deviation parametrization.}
\label{tab:priors-RD}
\small
\begin{tabular}{lrrrr}
Topic & Comparisons $\mu$ & Comparisons $\tau$ & Prior $\mu$ & Prior $\tau$ \\
\toprule
Acute Respiratory Infections & 308 (5857) & 152 (3184) & Student-t(0, 0.01, 1) & $\text{Normal}_+$(0, 0.10) \\ 
  Airways & 458 (8950) & 175 (3107) & Student-t(0, 0.01, 1) & $\text{Normal}_+$(0, 0.10) \\ 
  Anaesthesia & 413 (9585) & 234 (5983) & Student-t(0, 0.02, 1) & $\text{Normal}_+$(0, 0.11) \\ 
  Back and Neck & 14 (295) & 10 (212) & Student-t(0, 0.06, 2) & $\text{Normal}_+$(0, 0.10) \\ 
  Bone, Joint and Muscle Trauma & 206 (3781) & 86 (1476) & Student-t(0, 0.01, 1) & $\text{Normal}_+$(0, 0.12) \\ 
  Breast Cancer & 137 (2646) & 105 (2143) & Student-t(0, 0.05, 2) & $\text{Normal}_+$(0, 0.12) \\ 
  Childhood Cancer & 1 (13) & --- & Student-t(0, 0.02, 1) & --- \\ 
  Colorectal & 238 (4266) & 132 (2417) & Student-t(0, 0.01, 1) & $\text{Normal}_+$(0, 0.10) \\ 
  Common Mental Disorders & 650 (13334) & 455 (9880) & Student-t(0, 0.05, 1) & $\text{Normal}_+$(0, 0.11) \\ 
  Consumers and Communication & 44 (918) & 42 (895) & Student-t(0, 0.06, 2) & $\text{Normal}_+$(0, 0.10) \\ 
  Cystic Fibrosis and Genetic Disorders & 82 (1858) & 31 (786) & Student-t(0, 0.01, 1) & $\text{Normal}_+$(0, 0.09) \\ 
  Dementia and Cognitive Improvement & 126 (2086) & 67 (1038) & Student-t(0, 0.02, 1) & $\text{Normal}_+$(0, 0.09) \\ 
  Developmental, Psychosocial and Learning Problems & 107 (1467) & 96 (1322) & Student-t(0, 0.08, 2) & $\text{Normal}_+$(0, 0.13) \\ 
  Drugs and Alcohol & 107 (2065) & 70 (1445) & Student-t(0, 0.02, 1) & $\text{Normal}_+$(0, 0.10) \\ 
  Effective Practice and Organisation of Care & 59 (1106) & 45 (892) & Student-t(0, 0.06, 2) & $\text{Normal}_+$(0, 0.11) \\ 
  Emergency and Critical Care & 233 (4803) & 144 (3207) & Student-t(0, 0.02, 1) & $\text{Normal}_+$(0, 0.07) \\ 
  ENT & 17 (243) & 9 (116) & Student-t(0, 0.04, 1) & $\text{Normal}_+$(0, 0.12) \\ 
  Epilepsy & 117 (2085) & 67 (1355) & Student-t(0, 0.05, 2) & $\text{Normal}_+$(0, 0.09) \\ 
  Eyes and Vision & 58 (1017) & 38 (666) & Student-t(0, 0.04, 1) & $\text{Normal}_+$(0, 0.11) \\ 
  Fertility Regulation & 62 (1072) & 34 (562) & Student-t(0, 0.01, 1) & $\text{Normal}_+$(0, 0.08) \\ 
  Gut & 313 (6032) & 193 (3834) & Student-t(0, 0.05, 2) & $\text{Normal}_+$(0, 0.09) \\ 
  Gynaecological, Neuro-oncology and Orphan Cancer & 244 (5799) & 152 (3988) & Student-t(0, 0.02, 1) & $\text{Normal}_+$(0, 0.10) \\ 
  Gynaecology and Fertility & 375 (7017) & 204 (3929) & Student-t(0, 0.02, 1) & $\text{Normal}_+$(0, 0.09) \\ 
  Haematology & 155 (4546) & 82 (2740) & Student-t(0, 0.05, 1) & $\text{Normal}_+$(0, 0.09) \\ 
  Heart & 974 (20489) & 343 (6737) & Student-t(0, 0.00, 1) & $\text{Normal}_+$(0, 0.07) \\ 
  Heart; Vascular & 8 (178) & 8 (178) & Student-t(0, 0.08, 1) & $\text{Normal}_+$(0, 0.12) \\ 
  Hepato-Biliary & 1051 (36811) & 498 (14308) & Student-t(0, 0.01, 1) & $\text{Normal}_+$(0, 0.09) \\ 
  HIV/AIDS & 34 (501) & 17 (227) & Student-t(0, 0.02, 2) & $\text{Normal}_+$(0, 0.10) \\ 
  Hypertension & 66 (1127) & 13 (218) & Student-t(0, 0.01, 2) & $\text{Normal}_+$(0, 0.06) \\ 
  Incontinence & 78 (1479) & 61 (1183) & Student-t(0, 0.06, 2) & $\text{Normal}_+$(0, 0.12) \\ 
  Infectious Diseases & 361 (6614) & 228 (4141) & Student-t(0, 0.02, 1) & $\text{Normal}_+$(0, 0.10) \\ 
  Injuries & 217 (5692) & 128 (3529) & Student-t(0, 0.01, 0) & $\text{Normal}_+$(0, 0.14) \\ 
  Kidney and Transplant & 481 (8837) & 281 (5436) & Student-t(0, 0.03, 1) & $\text{Normal}_+$(0, 0.09) \\ 
  Lung Cancer & 76 (1416) & 65 (1284) & Student-t(0, 0.06, 2) & $\text{Normal}_+$(0, 0.12) \\ 
  Metabolic and Endocrine Disorders & 131 (3251) & 56 (1277) & Student-t(0, 0.00, 0) & $\text{Normal}_+$(0, 0.09) \\ 
  Methodology & 74 (2098) & 73 (2084) & Student-t(0, 0.10, 2) & $\text{Normal}_+$(0, 0.10) \\ 
  Movement Disorders & 59 (1058) & 48 (906) & Student-t(0, 0.06, 2) & $\text{Normal}_+$(0, 0.07) \\ 
  Multiple Sclerosis and Rare Diseases of the CNS & 29 (564) & 24 (484) & Student-t(0, 0.07, 2) & $\text{Normal}_+$(0, 0.10) \\ 
  Musculoskeletal & 139 (2403) & 80 (1269) & Student-t(0, 0.01, 1) & $\text{Normal}_+$(0, 0.11) \\ 
  Neonatal & 339 (6383) & 163 (2667) & Student-t(0, 0.02, 1) & $\text{Normal}_+$(0, 0.09) \\ 
  Neuromuscular & 43 (739) & 23 (395) & Student-t(0, 0.03, 1) & $\text{Normal}_+$(0, 0.08) \\ 
  Oral Health & 68 (1055) & 39 (541) & Student-t(0, 0.03, 1) & $\text{Normal}_+$(0, 0.12) \\ 
  Pain, Palliative and Supportive Care & 312 (6018) & 229 (4148) & Student-t(0, 0.11, 2) & $\text{Normal}_+$(0, 0.11) \\ 
  Pregnancy and Childbirth & 1310 (24437) & 724 (14321) & Student-t(0, 0.01, 1) & $\text{Normal}_+$(0, 0.08) \\ 
  Schizophrenia & 614 (13574) & 400 (9463) & Student-t(0, 0.03, 1) & $\text{Normal}_+$(0, 0.10) \\ 
  Sexually Transmitted Infections & 25 (318) & 16 (199) & Student-t(0, 0.04, 1) & $\text{Normal}_+$(0, 0.12) \\ 
  Skin & 145 (2347) & 99 (1580) & Student-t(0, 0.03, 1) & $\text{Normal}_+$(0, 0.12) \\ 
  Stroke & 303 (5390) & 132 (2230) & Student-t(0, 0.01, 1) & $\text{Normal}_+$(0, 0.06) \\ 
  Tobacco Addiction & 216 (5031) & 174 (4324) & Student-t(0, 0.04, 2) & $\text{Normal}_+$(0, 0.07) \\ 
  Urology & 91 (1728) & 67 (1357) & Student-t(0, 0.04, 1) & $\text{Normal}_+$(0, 0.10) \\ 
  Vascular & 185 (2570) & 85 (1224) & Student-t(0, 0.01, 1) & $\text{Normal}_+$(0, 0.12) \\ 
  Work & 14 (208) & 9 (144) & Student-t(0, 0.06, 2) & $\text{Normal}_+$(0, 0.12) \\ 
  Wounds & 75 (1169) & 43 (702) & Student-t(0, 0.01, 1) & $\text{Normal}_+$(0, 0.09) \\ 
  \bottomrule
  Pooled Estimate & 12042 (254326) & 6749 (141733) & Student-t(0, 0.03, 1) & $\text{Normal}_+$(0, 0.10) \\ 
\bottomrule
\end{tabular}
\end{table}

\end{document}